\documentclass[prd,a4paper,nofootinbib,showpacs,final,twocolumn,superscriptaddress,floatfix]{revtex4-1}

\usepackage{amsmath}
\usepackage{amssymb}
\usepackage{graphicx}
\usepackage[usenames,dvips]{color}
\usepackage[english]{babel}
\usepackage[latin1]{inputenc}
\usepackage{amsfonts}
\usepackage{appendix}
\usepackage{epsfig}
\usepackage{graphics,rotating}
\usepackage{dcolumn}
\usepackage{bm}
\usepackage{epstopdf}
\usepackage{color}
\usepackage[usenames,dvipsnames,svgnames]{xcolor}
\usepackage[colorlinks=true,
            linkcolor=blue,
            urlcolor=black,
            citecolor=magenta]{hyperref}
\usepackage{hyperref}
\usepackage[T1]{fontenc}
\usepackage{multirow}
\usepackage{float}
\usepackage{subfigure}
\usepackage{enumitem}

\begin{document}

\title{Spin supplementary conditions for spinning compact binaries}
\author{Bal\'{a}zs Mik\'{o}czi}
\date{\today }
\email{mikoczi.balazs@wigner.mta.hu} \affiliation{Research Institute
for Particle and Nuclear Physics, Wigner RCP H-1525 Budapest 114,
P.O.~Box 49, Hungary}

\begin{abstract}
We consider different \textit{spin supplementary conditions} (SSC)
for a spinning compact binary with the leading-order spin-orbit (SO)
interaction. The Lagrangian of the binary system can be constructed
but it is acceleration-dependent in two cases of SSC. We rewrite the
generalized Hamiltonian formalism proposed by Ostrogradsky and
compute the conserved quantities and the dissipative part of
relative motion during the gravitational radiation of each SSC. We
give the orbital elements and observed quantities of the SO
dynamics, for instance the energy and the orbital angular momentum
losses and waveforms and discuss their SSC dependence.
\end{abstract}

\maketitle

\section{Introduction}

The first direct observation of a gravitational-wave signal from two
coalescing black holes took place on September $14$, 2015
\cite{firstGW}. It has been confirmed that these compact binaries
are a main source of gravitational waves. The rough estimates also
indicate that typical stellar black hole ($\sim 30M_{\odot }$)
binaries can be observed $2-32$ times per year \cite{NTN}. Precise
measurements allow the observation of the finite size effects of
individual bodies, such as the masses and spins, with high accuracy.
The spin effects of these components can help the understanding of
several astrophysical processes, e.g., the \textit{spin-flip
phenomenon} \cite{ACST,GergelyBiermann}, \textit{frame dragging}
\cite{PW} and the evolution of \textit{accretion disks} around black
holes \cite{accr}.

The simple Lagrangian formalism of a relativistic spinning point
particle depends on the acceleration as demonstrated in Refs.
\cite{Riewe,Ellis}. The description of such a system is not unique
in generalized mechanics. The generalized Lagrangian formalism was
first developed by Jacobi and Ostrogradsky \cite{Ostro} in the 19th
century, see reviews Refs
\cite{Rodrigues,Whittaker,Nesterenko,ping}.

The description of spinning masses had been studied in general
relativity, but the first important result was achieved by Mathisson
\cite{Mathisson} who described the motion of extended bodies in
general cases, and he also generalized the mechanics of test bodies
in curved backgrounds in $1937$. In the early $1950$s Papapetrou
found the same equations of motion in Ref. \cite{Papapetrou}, but
his results came from a noncovariant formalism. Later the equations
of spinning bodies were improved in Refs. \cite{Tulczyjew,Dixon} and
since then they are called the
\textit{Mathisson-Papapetrou-Tulczyjew-Dixon equations} (afterwards
MPTD-equations).

It is well known that this system of equations is not closed;
therefore we have to impose some \textit{spin supplementary
conditions} (afterwards SSC). In the literature there are basically
four SSCs, namely the \textit{Frenkel-Mathisson-Pirani}, the
\textit{Newton-Wigner-Pryce}, the \textit{Corinaldesi-Papapetrou},
and the \textit{Tulczyjew-Dixon} (see, Refs. \cite{Frenkel,
Mathisson,Pirani,Papapetrou,CorPap,NW,Pryce,Tulczyjew,Dixon,Moller,Ohashi,KS}).
The MPTD-equations have been used with these SSCs to study the
motion of a test spinning particle in different curved backgrounds
and in the ultra-relativistic regime (see
Refs.\cite{Plyatsko2013,Puetzenfeld,LukesGerakopoulos,Harms2016,Deriglazov2015a,Deriglazov2015b}).

The first effective description of the leading-order spin effect in
the \textit{post-Newtonian} sense (hereafter PN) was given by
Tulczyjew in Ref. \cite{TulczyjewSO}. The nongeodesic motion of test
particles and compact binaries with PN corrections were first
developed by Barker and O'Connell for different SSCs in
\cite{BOC74,BOC75}. The acceleration of the compact binary with
leading-order \textit{spin-orbit interaction} (hereafter SO) is not
obvious, but it rather depends on the chosen SSC \cite{Kidder}. Some
authors have first investigated the spin effects with the help of
the PN approach for some of the SSCs in Refs.
\cite{DS,KWW,Kidder,Wex,Kepler,GPV3}. The Lagrangian of compact
binaries with SO interaction is \textit{acceleration-dependent} in
two cases of SSC \cite{KWW,Kepler}, but the Lagrangian does not
depend on acceleration terms for the \textit{Newton-Wigner-Pryce
SSC} in Refs. \cite{DS,Wex,Gopa}. Taking into account the spins of
the bodies in physical systems leads to additional extra degrees of
freedom, for instance, \textit{spin-precession equations}, which are
important for the investigation of classical and/or quantum systems
\cite{ACST}. It is important to know how the motion of the spins
changes the orbital evolution and the dissipation under
gravitational radiation for compact sources.

Recently, a simple Hamiltonian in ADM coordinates for covariant SSC
has been found by \cite{DJS}; it follows the leading-order and
next-to-leading-order equations of motion of SO contribution in
harmonic coordinates in \cite{TOO,FBB}. Recently, \textit{effective
field theory} (EFT) methods have been used for the computation of
the spin-orbit, spin-spin and self-spin contributions in
\cite{Porto2006,Porto2010,Levi2010}. Nowadays, there are also EFT
results for the 4PN-order spin contributions
\cite{LS01,LS02,LS03,LS}.

In this paper we give the Lagrangian of the compact spinning binary
with leading-order SO interaction for well-known SSCs. We calculate
all the conserved quantities and orbital parameters for these SSCs.
As the main result we give the classical relative orbital evolution
of a spinning binary system. Because the Lagrangian is
acceleration-dependent in two SSCs, we compute the generalized
Lagrangian and its canonical dynamics. We construct the nontrivial
type of the Ostrogradsky Hamiltonian method, and then we show two
examples for the elimination of acceleration dependent terms from
Lagrangians, i.e., the constrained dynamics in Refs.
\cite{Riewe,Ellis} and the double zero method proposed by Barker and
O'Connell in Refs. \cite{BOC80,BOC80b,BOC80c,BOC86}.

Moreover, we consider the dissipative part of the evolution of the
compact binary. We calculate the symmetric trace-free
(STF)-multipole moments and investigate the instantaneous energy and
the orbital angular momentum losses. It can be seen that these
instantaneous losses depend on SSC, but the SSC dependence
disappears after averaging over one orbital period. Finally, we
compute the gravitational waveforms for all SSCs, where it can be
seen that the leading-order contribution is independent of SSC, but
the next-to-leading-order terms depend on SSC.

Our Lagrangian (and Hamiltonian) formalism is consistent with the
equations of motion for all SSCs in Ref. \cite{Kidder}. Several
authors eliminated the covariant SSC at the level of the potential
using the Dirac bracket method and the variation of the action
principle in Refs. \cite{Levi2010,HSS}. They first applied the
nonreduced SO part of the potential and achieved the reduced
Hamiltonian form of dynamics, and their result is consistent with
one of the Barker O'Connell type equations of motion if we use the
baryonic coordinate transformation (see Ref. \cite{Porto2006}).

This article is organized as follows. In Sec. II we introduce the
MPTD-equations and then we focus on the SSCs. In Sec. III we discuss
the generalized Lagrangian and canonical mechanics of the compact
binary system with spin-orbit interaction for all SSCs. We construct
the canonical (Ostrogradsky type) dynamics and we demonstrate the
elimination of the acceleration-dependent terms from the Lagrangian
using the constrained dynamics and the double zero method in
Appendix A and Appendix B, respectively. We rewrite the equations of
motion from Lagrangian formalism in Sec. IV, and then we add the
radial and angular motion in a simple case in Sec. V. In Sec. VI we
compute the energy and the angular momentum losses due to
gravitational radiation, and finally we calculate the waveform of
the SO interaction terms for all SSCs in Sec. VII. At the end of
this paper Appendices A-C describe the relationship between the
Hamiltonian and Lagrangian formalisms.

In this paper Greek indices $\alpha $, $\beta $,... run from $0$ to
$3$ and the Roman indices $a$, $b$,... run from 1 to 3. The repeated
Greek (Roman) indices in a row mean Einstein's summation from $1$
($0$) to $3$. Generally, we use lowercase indices for spatial
tensors. We use angular and square brackets for symmetrized and
antisymmetrized indices, respectively, e.g.,
$T_{(ab)}=(T_{ab}+T_{ba})/2$ and $T_{[ab]}=(T_{ab}-T_{ba})/2$. The
fully symmetric trace-free part of tensor will be denoted by "STF,"
$(T_{ab})^{STF}\equiv T_{<ab>}=T_{(ab)}-\delta _{ab}T_{(cc)}/3$. The
use of the transverse-traceless part of the tensor will be denoted
by "TT," $(T_{ab})_{TT}=\Lambda _{ab,cd}T_{cd}$ where $\Lambda
_{ab,cd}=P_{ai}P_{jb}-P_{ab}P_{cd}/2$ and $P_{ab}=\delta
_{ab}-N_{a}N_{b}$ is the projector and the vector $\mathbf{N}$ is
the line of sight. The $G$ is the gravitational constant and $c$ is
the speed of light. Each calculation of the spin-orbit coupling is
valid only to the leading-order contributions with a $1.5$PN
accuracy.

\section{Spin supplementary conditions}

The MPTD-equations of motion of the spinning body in general
relativity in Refs. \cite{Mathisson,Papapetrou} are
\begin{eqnarray}
\frac{DS^{\alpha \beta }}{D\tau } &=&p^{\alpha }u^{\beta }-p^{\beta
}u^{\alpha }\text{ },  \label{Pap1} \\
\frac{Dp^{a}}{D\tau } &=&-\frac{1}{2}R_{\gamma \delta \beta }^{\alpha
}S^{\delta \beta }u^{\gamma }\text{ },  \label{Pap2}
\end{eqnarray}
where $\tau $ is the affine parameter of the trajectory, $p^{\alpha
}$ is the four-momentum, $u^{\alpha }=dx^{\alpha }/d\tau$ is the
tangent vector to the trajectory, $S^{\alpha \beta }$ is the skew
canonical spin tensor which represents the internal angular
three-momentum after using some SSC, i.e., \textit{spin}, $R_{\beta
\gamma \delta }^{\alpha }$ is the Riemann tensor and $D/D\tau
=u^{\alpha }\nabla _{\alpha }$ is the covariant derivative along
$u^{\alpha }$. The spin vector is given by $S_{\alpha }=-\varepsilon
_{\alpha \beta \mu \nu }u^{\beta }S^{\mu \nu }/2$ for Frenkel-
Mathisson-Pirani SSC where $\varepsilon _{\alpha \beta \mu \nu }$ is
the four-dimensional Levi-Civita tensor (we use the unit $c=1$ in
this section). The three-dimensional spin vector can be obtained
from the use of any SSC (see \cite{Wex}). We assume the $u^{\alpha
}u_{\alpha }=-1$ for the four-velocity. There are some scalars,
i.e., the rest mass $m=-p^{\alpha }u_{\alpha }$ with respect to
$u^{\alpha }$, the "other" rest mass $\mu^{2}=-p^{\alpha }p_{\alpha
}$ with respect to $p^{\alpha }$, and the magnitude of spin
$2s^{2}=S^{\alpha \beta }S_{\alpha \beta }$. These quantities are
not conserved for all cases, e.g. the $m$ is conserved for the
Frenkel-Mathisson-Pirani SSC, the $\mu$ is conserved for the
Tulczyjew-Dixon SSC, and the $s$ is conserved for the
Tulczyjew-Dixon and the Frenkel-Mathisson-Pirani SSCs (see details,
e.g., Ref. \cite{LukesGerakopoulos}). Thus, the variables of the
Eqs. (\ref{Pap1}) and (\ref{Pap2}) are more than the number of
equations, so we have to impose the spin supplementary condition. In
the literature, there are basically four SSCs: the
\textit{Frenkel-Mathisson-Pirani} (hereafter SSC I)
\cite{Frenkel,Mathisson,Pirani}, the \textit{Newton-Wigner-Pryce}
(SSC II) \cite{NW,Pryce}, the \textit{Corinaldesi-Papapetrou} (SSC
III) \cite{CorPap} and the \textit{Tulczyjew-Dixon} (SSC IV)
\cite{Tulczyjew,Dixon,Moller}.
\begin{eqnarray}
\text{ }S^{\alpha \beta }u_{a} &=&0\text{ }\qquad \text{SSC I ,} \\
2S^{0\beta }+u_{\alpha }S^{\alpha \beta } &=&0\text{ }\qquad \text{SSC II ,}
\\
S^{\alpha 0} &=&0\text{ }\qquad \text{SSC III ,} \\
S^{\alpha \beta }p_{\alpha } &=&0\text{ }\qquad \text{SSC IV .}
\end{eqnarray}
First, SSC I appeared in the description of the spin of electrons in
\cite{Frenkel}. This condition is also called the \textit{covariant
SSC}. In this SSC Weyssenhoff and Raabe pointed out the appearance
of the helical motion which is unphysical \cite{WR}. However,
recently this motion was interpreted by a hidden
electromagnetic-like momentum \cite{CHNZ}. SSC II was first used for
quantum mechanics because it is well known that the center of mass
of a rotating particle is not invariant under the Lorentz
transformation, see Refs. \cite{NW,Pryce}. This SSC II has been
generalized for curved spacetimes in \cite{BRB}. Our definition of
SSC II is equivalent to Eqs. (4.6) and (4.7) in Ref. \cite{BRB} for
flat spacetime where the unit timelike vector field reduce to
Kronecker delta. The simplest way is to choose SSC III where the
timelike components were dropped by Corinaldesi and Papapetrou in
Ref. \cite{CorPap}. Barker and O'Connell have found that the
macroscopic limit of the potential of two quantum spinning masses
with spin $1/2$ from the quantum theory of gravitation by Gupta,
which leads to the acceleration of SSC II in Refs.
\cite{Gupta,BOC75}. SSC I and SSC IV are equivalents of each other
if we neglect the quadratic terms in spin. This should be valid for
the spin-orbit interaction because this interaction is linear in
spin. The transformations between the SSCs were described by Ref.
\cite{BOC74} for spinning test particles of the nongeodesic motion.
We note that the Lagrangian of the spin-orbit interaction of compact
binaries does not depend on the acceleration only in SSC II, and the
acceleration-dependent terms appear in other SSCs \cite{Kepler}.
Note that recently another SSC has been given by Ohashi, Kyrian and
Semer\'{a}k in Refs. \cite{Ohashi,KS}. For more details see Ref.
\cite{CN}.

\begin{figure}[!hb]
\includegraphics[width=0.9\columnwidth]{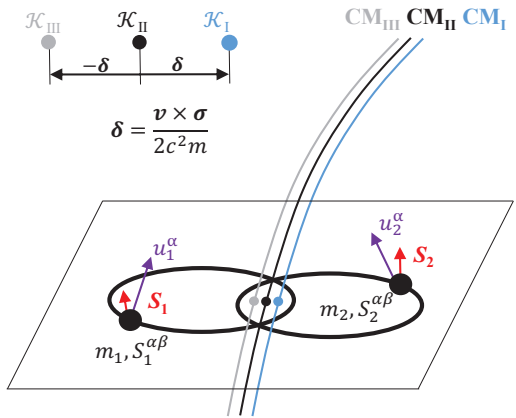}
\caption{Different world lines of the center of spinning masses.
Three centers of mass (CMs) have been demonstrated for each SSC.}
\label{fig1}
\end{figure}

\section{Generalized mechanics of spinning two-body systems}

\label{SO} Consider a compact two-body system with leading-order
spin-orbit interaction, where masses are $m_{i}$ and spins are
$\mathbf{S}_{\mathbf{i}}$ ($i=1,2$). The equation of motion
(relative acceleration) for the three different SSCs can be found in
Ref. \cite{Kidder} as
\begin{eqnarray}
\mathbf{a}
&=&-\frac{Gm}{r^{3}}\mathbf{r+}{\frac{G}{c^{2}r^{3}}}\Biggl\{{\frac{3}{r^{2}}}\mathbf{r}\left[
(\mathbf{r}\times \mathbf{v})\cdot \left(
2\mathbf{S}+(1+k)\mbox{\boldmath
$\sigma$}\right) \right]  \notag \\
&&-\mathbf{v\times (}4\mathbf{S}+3{\mbox{\boldmath
$\sigma$})}+{\frac{3\dot{r}}{r}}\mathbf{r}\times
(2\mathbf{S}+(2-k){\mbox{\boldmath $\sigma$}})\Biggr\},  \label{acc}
\end{eqnarray}
where $r=\left\vert \mathbf{r}\right\vert $ and $\mathbf{v}$ are the
relative distance and velocity, respectively, $m=m_{1}+m_{2}$ is the
total mass of the system where the masses $m_{1}$ and $m_{2}$, and
the overdot denotes the derivative with respect to the time,
$\mathbf{S}=\mathbf{S}_{1}+\mathbf{S}_{2}$ is the total spin vector
and ${\mbox{\boldmath
$\sigma$}}=(m_{2}/m_{1})\mathbf{S}_{1}+(m_{1}/m_{2})\mathbf{S}_{2}$
is the weighted spin vector where the individual spins
$\mathbf{S}_{1}$ and $\mathbf{S}_{2}$ follow the notations of Ref.
\cite{GPV3}. Here we have introduced the SSC-dependent parameter $k$
with the following values for the different SSCs
\begin{equation}
k=\left\{
\begin{array}{ll}
1 & \text{for SSC I, } \\
{\frac12}
& \text{for SSC II, } \\
0 & \text{for SSC III. }
\end{array}
\right.
\end{equation}
The transformation between the SSCs (see Fig. \ref{fig1}) is given
by Ref. \cite{Kidder} as
\begin{eqnarray}
\mathbf{r}^{(k)} &=&\mathbf{r}^{(k\prime )}+\frac{k^{\prime
}-k}{c^{2}m}\mathbf{v}\times {\mbox{\boldmath
$\sigma$}}\ , \\
\mathbf{v}^{(k)} &=&\mathbf{v}^{(k\prime )}+\frac{G(k-k^{\prime
})}{c^{2}r^{3}}\mathbf{r}\times {\mbox{\boldmath $\sigma$}}\ .
\end{eqnarray}
Then we can compute the corresponding Lagrangian from the
acceleration in Ref. \cite{KWW} for SSC I and Ref. \cite{Kepler} for
SSC II, as
\begin{eqnarray}
\mathcal{L} &=&\frac{\mu }{2}\mathbf{v}^{2}+\frac{Gm\mu
}{r}+\frac{G\mu } {c^{2}r^{3}}\mathbf{v\cdot }\left[
\mathbf{r}\times \left( 2\mathbf{S}+(1+k)
{\mbox{\boldmath $\sigma$}}\right) \right]  \notag \\
&&+\frac{(2k-1)\mu }{2c^{2}m}\mathbf{v}\cdot (\mathbf{a}\times
{\mbox{\boldmath $\sigma$})},  \label{LagrangeSO}
\end{eqnarray}
where $\mu =m_{1}m_{2}/m$ is the reduced mass. It can be seen that
the only case in which the Lagrangian does not depend on
acceleration terms is that of SSC II (for $k=1/2$). According to
Ref. \cite{S84} the infinitesimal acceleration dependence can be
eliminated by a time-coordinate transformation. Here only the
$k=1/2$ case is relevant, but in this way the SSC dependence is
shifted in the coordinates. Note that the acceleration dependence
can be eliminated if we use the Newtonian-order acceleration
$\mathbf{a}_{N}=-(Gm/r^{3})\mathbf{r}$ in Eq.
(\ref{LagrangeSO})\thinspace; thus, we get the case of SSC II. The
generalized moments can be calculated by the generalized Lagrangian
as
\begin{equation}
\mathbf{p}=\frac{\partial \mathcal{L}}{\partial
\mathbf{\dot{r}}}-\mathbf{\dot{q}},\qquad \mathbf{q}=\frac{\partial
\mathcal{L}}{\partial \mathbf{a}},
\end{equation}
yields to
\begin{equation}
\mathbf{p}=\mu \mathbf{\dot{r}}+\frac{G\mu
}{c^{2}r^{3}}\mathbf{r}\times \left[
2\mathbf{S}+(2-k){\mbox{\boldmath $\sigma$}}\right] ,\qquad
\mathbf{q}=\frac{(2k-1)\mu }{2c^{2}m}{\mbox{\boldmath
$\sigma$}}\times \mathbf{\dot{r}}.  \label{Lageq}
\end{equation}
The energy and the orbital angular momentum from
acceleration-dependent Lagrangian dynamics are respectively
\begin{eqnarray}
E &=&\mathbf{p}\cdot \mathbf{v+q}\cdot \mathbf{a-}\mathcal{L},
\label{energy} \\
\mathbf{L} &=&\mathbf{r}\times \mathbf{p+v}\times \mathbf{q}.
\label{momentum}
\end{eqnarray}
The energy $E$, the magnitude of the orbital angular momentum
$L=\left\vert \mathbf{L}\right\vert $, the magnitudes of the
spinvectors $S_{i}$ and the total angular momentum vector
$\mathbf{J}=\mathbf{L}+\mathbf{S}$ are conserved quantities. We
compute the energy and the orbital angular momentum for different
SSCs using Eqs. (\ref{energy}) and (\ref{momentum}),
\begin{eqnarray}
E &=&\frac{\mu }{2}\mathbf{v}^{2}-\frac{Gm\mu }{r}+\frac{G\mu (1-2k)}
{c^{2}r^{3}}\mathbf{v}\cdot \left( \mathbf{r}\times {\mbox{\boldmath $
\sigma$}}\right) ,  \label{energy_exp} \\
\mathbf{L} &=&\mu \mathbf{r}\times \mathbf{v}+\frac{G\mu
}{c^{2}r^{3}}\mathbf{r}\times \left[ \mathbf{r}\times \left(
2\mathbf{S}+(2-k){\mbox{\boldmath
$\sigma$}}\right) \right]  \notag \\
&&+\frac{(1-2k)\mu }{2c^{2}m}\mathbf{v}\times \left(
\mathbf{v}\times {\mbox{\boldmath $\sigma$}}\right) .
\label{momentum_exp}
\end{eqnarray}
Here the main conserved quantities, i.e., the energy $E$ and the
magnitude of orbital angular momentum $L$, depend on SSC although we
do not mark the SSC dependence ($k$ dependence) on $E$ and $L$. If
we set $k=1/2$ the Lagrangian does not depend on the acceleration.

Considering the canonical dynamics the first Hamiltonian description
of the spin-orbit interaction for compact binary systems in SSC II
was given by Refs. \cite{Wex},\cite{Gopa}, and \cite{Tessmer}, we
can calculate the Hamiltonian from (acceleration-dependent)
Lagrangian for all SSCs. The generalized Hamiltonian from a
generalized Legendre transformation is
\begin{equation}
\mathcal{H}=\mathbf{p}\cdot \mathbf{v}+\mathbf{q}\cdot
\mathbf{a}-\mathcal{L}.  \label{Legendre}
\end{equation}
Here we should eliminate the acceleration terms from the
Hamiltonian, Eq. (\ref{Legendre}); thus, we need to use the
acceleration in Eq. (\ref{acc}). Two canonical pairs appear here,
which are $(\mathbf{r},\mathbf{p})$ and $(\mathbf{v},\mathbf{q})$.
This is nontrivial because we do not know which canonical moment to
use in the Legendre transformation. After using the canonical moment
$\mathbf{p}$ in Eq. (\ref{Lageq}), we can calculate the
Hamiltonian\footnote{If we do not use the canonical moments, but
just straightforwardly keep the first and second terms
$\mathbf{p}\cdot \mathbf{v}$, $\mathbf{q}\cdot \mathbf{a}$ and
eliminate the acceleration $\mathcal{L}(\mathbf{r,v,a})\rightarrow
\mathcal{L}(\mathbf{r,v})$, then we get
$\mathcal{H}_{L}=\mathbf{-}\frac{\mu }{2}\mathbf{v}^{2}-\frac{Gm\mu
}{r}+\mathbf{p}\cdot \mathbf{v+q}\cdot \mathbf{a}-\frac{G\mu
}{2c^{2}r^{3}}\mathbf{v\cdot }\left[ \mathbf{r}\times \left(
4\mathbf{S}+3{\mbox{\boldmath $\sigma$}}\right) \right]$, where
$\mathbf{a=a(r,v)}$ is the acceleration from the Lagrangian. This
Hamiltonian satisfies the generalized Hamilton's Eqs. (\ref{Heq1})
and (\ref{Heq2}), but it is not consistent for Newtonian limit by
$\mathcal{H}=\mathbf{p}^{2}/(2\mu )-Gm\mu /r$.}
\begin{eqnarray}
\mathcal{H} &=&\frac{\mathbf{p}^{2}}{2\mu }-\frac{Gm\mu }{r}  \notag \\
&&+\frac{G}{2c^{2}r^{3}}\mathbf{r}\cdot \biggl [2\mathbf{p}\times
\left( 2\mathbf{S}+(2-k){\mbox{\boldmath
$\sigma$}}\right)   \notag \\
&&+(2k-1)\mu \mathbf{v}\times {\mbox{\boldmath
$\sigma$}}\biggr ]-\frac{Gm}{r^{3}}\mathbf{q}\cdot \mathbf{r}  \notag \\
&&\mathbf{-}{\frac{G}{c^{2}\mu
r^{3}}}\mathbf{q}\cdot\Biggl\{\mathbf{p\times
(}4\mathbf{S}+3{\mbox{\boldmath
$\sigma$})}  \notag \\
&&-\mathbf{r}\left[ (\mathbf{r}\times \mathbf{p})\cdot
(2\mathbf{S}+(1+k)\mbox{\boldmath
$\sigma$})\right]   \notag \\
&&-{\frac{3(\mathbf{r}\cdot \mathbf{p)}}{r^{2}}}\mathbf{r}\times
(2\mathbf{S}+(2-k){\mbox{\boldmath $\sigma$}})\Biggr\}.
\end{eqnarray}
Note that we had to add an extra term $(1-2k)G/(2c^{2}r^{3})\left(
\mu \mathbf{v-p}\right) \cdot (\mathbf{r}\times {\mbox{\boldmath
$\sigma$})}$ (which disappears if we use the zeroth-order canonical
moment $\mathbf{p}$) to the original Hamiltonian in Eq.
(\ref{Legendre}); otherwise, it could not satisfy the generalized
Hamilton's equations
\begin{eqnarray}
\mathbf{\dot{p}} &=&-\frac{\partial \mathcal{H}}{\partial \mathbf{r}},\qquad
\mathbf{\dot{r}}\mathbf{=}\frac{\partial \mathcal{H}}{\partial \mathbf{p}},
\label{Heq1} \\
\mathbf{\dot{q}} &=&-\frac{\partial \mathcal{H}}{\partial \mathbf{v}},\qquad
\mathbf{\dot{v}=}\frac{\partial \mathcal{H}}{\partial \mathbf{q}}.
\label{Heq2}
\end{eqnarray}
Then the explicit Hamilton's equations up to $\mathcal{O}(c^{-2})$
are
\begin{eqnarray}
\mathbf{\dot{p}} &=&-\frac{Gm\mu
}{r^{3}}\mathbf{r}-\frac{G}{c^{2}r^{3}}\mathbf{p}\times \left[
2\mathbf{S}+(1+k){\mbox{\boldmath
$\sigma$}}\right]   \notag \\
&&+\frac{3G}{c^{2}r^{5}}\mathbf{r}\left[ (\mathbf{r}\times
\mathbf{p})\cdot
(2\mathbf{S}+(1+k)\mbox{\boldmath $\sigma$})\right] ,  \label{Ham1} \\
\mathbf{\dot{r}} &\mathbf{=}&\frac{\mathbf{p}}{\mu
}-\frac{G}{c^{2}r^{3}}\mathbf{r}\times \left[
2\mathbf{S}+(2-k){\mbox{\boldmath
$\sigma$}}\right] ,  \label{Ham2} \\
\mathbf{\dot{q}} &\mathbf{=}&\frac{G\mu
(2k-1)}{2c^{2}r^{3}}\mathbf{r}\times \mbox{\boldmath
$\sigma$}\mathbf{,}  \label{Ham3} \\
\mathbf{\dot{v}}
&\mathbf{=}&-\frac{Gm}{r^{3}}\mathbf{r-}{\frac{G}{c^{2}\mu
r^{3}}}\Biggl\{\mathbf{p\times (}4\mathbf{S}+3{\mbox{\boldmath
$\sigma$})}  \notag \\
&&-{\frac{3}{r^{2}}}\mathbf{r}\left[ (\mathbf{r}\times
\mathbf{p})\cdot (2\mathbf{S}+(1+k)\mbox{\boldmath
$\sigma$})\right]   \notag \\
&&-\frac{3(\mathbf{r}\cdot \mathbf{p)}}{r^{2}}\mathbf{r}\times
(2\mathbf{S}+(2-k){\mbox{\boldmath $\sigma$}})\Biggr\}, \label{Ham4}
\end{eqnarray}
where in Eqs. (\ref{Ham1}) and (\ref{Ham2}) we used the
approximation $\mathcal{O}(\mathbf{q})\mathcal{O}(c^{-2})\approx 0$,
which can be seen from Eq. (\ref{Lageq}) or Eq. (\ref{Ham3}).
Equation (\ref{Ham3}) disappears for SSC II and Eqs. (\ref{Ham1})
and (\ref{Ham2}) will be equivalent to Eq. (\ref{Ham4}) which is the
acceleration Eq. (\ref{acc}) in the Lagrangian method. In Appendices
A and B we show two different methods for the elimination of the
acceleration-dependent terms from the Lagrangian.

\subsection{Canonical structure}

We define the generalized Poisson brackets following the paper of
Ref. \cite{YH80}, where $f$ and $g$ functions arbitrarily depend on
the canonical and spin variables
\begin{equation}
\left\{ f,g\right\} =\sum_{j,i=1}^{3,2}\left[ \frac{\partial
f}{\partial \mathbf{Q}_{i}^{j}}\frac{\partial g}{\partial
\mathbf{P}_{i}^{j}}-\frac{\partial f}{\partial
\mathbf{P}_{i}^{j}}\frac{\partial g}{\partial
\mathbf{Q}_{i}^{j}}-\frac{\partial f}{\partial
\mathbf{S}_{i}^{j}}\left( \mathbf{S}_{i}^{j}\times \frac{\partial
g}{\partial \mathbf{S}_{i}^{j}}\right) \right] ,
\end{equation}
where $\mathbf{Q}_{1}\equiv \mathbf{r}$, $\mathbf{Q}_{2}\equiv
\mathbf{v} $, $\mathbf{P}_{1}\equiv \mathbf{p}$ and
$\mathbf{P}_{2}\equiv \mathbf{q}$ are useful vector notations, and
the superscripts are the components of the vectors. Thus, the
nonvanishing fundamental Poisson brackets are
\begin{eqnarray}
\left\{ P_{i}^{l},Q_{j}^{k}\right\}  &=&\delta _{ij}\delta _{lk},
\label{fund1} \\
\left\{ S_{i}^{l},S_{j}^{k}\right\}  &=&\delta _{ij}\varepsilon
_{lkm}S_{i}^{m}.  \label{fund2}
\end{eqnarray}
The time evolutions are given by their Poisson brackets with the
Hamiltonian, so the generalized Hamilton's equations can be written
as
\begin{eqnarray}
\mathbf{\dot{P}}_{i} &\mathbf{=}&\left\{ \mathbf{P}_{i}\mathbf{,}\mathcal{H}\right\} , \\
\mathbf{\dot{Q}}_{i} &\mathbf{=}&\left\{
\mathbf{Q}_{i}\mathbf{,}\mathcal{H}\right\} .
\end{eqnarray}
The time evolution of the spins, the orbital angular momentum, and
the Laplace-Runge-Lenz vector can be computed using of the
fundamental Poisson brackets Eqs. (\ref{fund1}) and (\ref{fund2})
\begin{eqnarray}
\mathbf{\dot{S}}_{i} &=&\left\{
\mathbf{S}_{i}\mathbf{,}\mathcal{H}\right\} =\frac{G(4+3\nu
_{i})}{2c^{2}r^{3}}\mathbf{L}\times \mathbf{S}_{i},
\label{Sdot} \\
\mathbf{\dot{L}} &=&\left\{ \mathbf{L,}\mathcal{H}\right\}
=\frac{G}{2c^{2}r^{3}}\left( 4\mathbf{S}+3{\mbox{\boldmath
$\sigma$}}\right) \times \mathbf{L,}  \label{Ldot} \\
\mathbf{\dot{A}} &=&\left\{ \mathbf{A,}\mathcal{H}\right\}
=\frac{G}{c^{2}r^{3}}\left[ 2\mathbf{S}+(2-k){\mbox{\boldmath
$\sigma$}}\right] \times \mathbf{A}  \notag \\
&&+\frac{3G}{c^{2}\mu r^{5}}\left( \mathbf{r}\times
\mathbf{L}\right) \left[ 2\mathbf{L}\cdot
\mathbf{S}+(1+k)\mathbf{L}\cdot {\mbox{\boldmath
$\sigma$}}\right]   \notag \\
&&+\frac{G(2k-1)}{c^{2}r^{3}}\left[ \frac{\mu
v^{2}}{2}{\mbox{\boldmath $\sigma$}}\times
\mathbf{r}+\mathbf{(L\cdot }\mbox{\boldmath
$\sigma$}\mathbf{)v}\right] .  \label{Adot}
\end{eqnarray}
with $\nu _{1}=\nu $ and $\nu _{2}=\nu ^{-1}$ as shorthand notations
where $\nu =$ $m_{2}/m_{1}$ is the mass ratio of the compact binary
and $\mathbf{A=}\frac{\mathbf{p}}{\mu }\times \mathbf{L-}\frac{Gm\mu
}{r}\mathbf{r}$ is the Laplace-Runge-Lenz (LRL) vector.\footnote{The
magnitude of the zeroth-order LRL ($A_{N}=\left\vert
\mathbf{A_{N}}\right\vert $) is conserved. The relationship between
$A_{N}$, $L_{N}$, and $E_{N}$ is $\mu A_{N}^{2}=G^{2}m^{2}\mu
^{3}+2E_{N}L_{N}^{2}$. The Newtonian geometric condition
$\mathbf{L}\cdot \mathbf{A}=0$, which contains the spin-orbit
contributions, is only satisfied for SSC II and for the single spin
limit of the spin-orbit interaction in Ref. \cite{GPV1}.} To derive
explicit evolution equations, we had to use the integration of the
equation $\mathbf{\dot{q}}\mathbf{=}(2k-1)\mu
/(2c^{2}m){\mbox{\boldmath $\sigma$}}\times \mathbf{\dot{v}}$ from
Eqs. (\ref{Ham3}) and (\ref{Ham4}). It can be seen that the time
evolution of the LRL vector depends on SSC and is not a pure
precession as in the case of $\mathbf{\dot{S}}_{i}$ and
$\mathbf{\dot{L}}$. \footnote{We can get a pure precession using the
Newtonian orbital average of Eq. (\ref{Adot}) only for SSC II (see
Refs \cite{BOC70} and \cite{DS88}).}

\section{The equations of motion}

\begin{figure}[!hb]
\includegraphics[width=0.5\columnwidth]{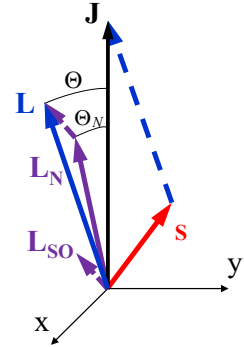}
\caption{Total and Newtonian angular momenta. The angles of $\Theta$
and $\Theta _{N}$ are different but the evolution equations for the
other two Euler angles are equivalent with each other.} \label{fig2}
\end{figure}

We need to compute the evolution of the angular momenta. The
evolution of the Newtonian orbital angular momentum vector
$\mathbf{L}_{\mathbf{N}}=\mu \mathbf{r}\times \mathbf{v}$ the first
term in Eq. (\ref{momentum_exp}) does not follow a pure precession
motion, since
\begin{eqnarray}
\mathbf{\dot{L}}_{\mathbf{N}} &=&-\frac{\mu
G}{c^{2}r^{3}}\mathbf{r}\times \left[ \mathbf{v\times
(}4\mathbf{S}+3{\mbox{\boldmath $\sigma$})}\right]
\notag \\
&&+\frac{3\mu G\dot{r}}{c^{2}r^{4}}\mathbf{r}\times \left[
\mathbf{r}\times (2\mathbf{S}+(2-k){\mbox{\boldmath
$\sigma$}})\right] ;  \label{Newtonian_ang}
\end{eqnarray}
meanwhile, the motion of the total orbital angular momentum vector
$\mathbf{L}$ leads to a pure precession equation from Eq.
(\ref{Ldot}),
\begin{equation}
\mathbf{\dot{L}}\mathbf{=}\frac{G}{2c^{2}r^{3}}\left(
4\mathbf{S}+3{\mbox{\boldmath $\sigma$}}\right) \times \mathbf{L}.
\label{total_ang}
\end{equation}
This pure precession can be given by the conservation of the total
angular momentum ($\mathbf{\dot{J}}=0$), and as a consequence
$\mathbf{\dot{L}=-\dot{S}}$. This way the motion of the total spin
vector does not have pure precession, but the individual spin
vectors of the orbiting bodies do have one in Eq. (\ref{Sdot}) or in
Ref. \cite{BOC74}. It can be seen that the pure precession equation
for the total angular momentum vector does not depend on SSC, but
the evolution of the Newtonian angular momentum vector depends on
SSC; see Eqs. (\ref{total_ang}) and (\ref{Newtonian_ang}),
respectively, and Fig \ref{fig2}. We may get different angular
equations depending on which orbital momentum vectors
($\mathbf{L}_{\mathbf{N}}$ or $\mathbf{L}$) we measure with the
Euler angles. The radial motion is invariant to this choice.
Hereafter, we only consider the orbital motion involved in dynamical
quantities fixed to $\mathbf{L}_{\mathbf{N}}$, where we will give
the full radial motion for each of the other SSCs and we neglect the
total angular motion (see Refs.
\cite{Wex,GPV3,Gopa,Racine,Tessmer,recoil}).

We compute the orbital evolution using the conserved quantities $E$
and $L$ in Eqs. (\ref{energy_exp}) and (\ref{momentum_exp}) for
different SSCs. The first integrals can be separated into radial and
angular motion. The radial and angular motion from energy and the
magnitude of the orbital angular momentum are governed by
\begin{eqnarray}
\dot{r}^{2} &=&\dot{r}_{N}^{2}-\frac{2G}{c^{2}\mu r^{3}}(2\mathbf{L}\cdot
\mathbf{S}+(2-k)\mathbf{L}\cdot {\mbox{\boldmath
$\sigma$})}  \notag \\
&&+\frac{2(2k-1)E}{c^{2}m\mu ^{2}r^{2}}(\mathbf{L}\cdot
{\mbox{\boldmath
$\sigma$}),}  \label{radial} \\
\dot{\varphi} &=&\dot{\varphi}_{N}+\frac{G}{c^{2}Lr^{3}}(2\mathbf{L}\cdot
\mathbf{S}+3(1-k)\mathbf{L}\cdot {\mbox{\boldmath
$\sigma$})}  \notag \\
&&{-}\frac{(2k-1)E}{c^{2}m\mu Lr^{2}}(\mathbf{L}\cdot
{\mbox{\boldmath $\sigma$}),}  \label{ang}
\end{eqnarray}
where $\varphi $ is the azimuthal angle on the orbital plane. We
have introduced the Newtonian formulas where $\dot{r}_{N}^{2}=2E\mu
^{-1}+2Gmr^{-1}-L^{2}r^{-2}\mu ^{-2}$ and $\dot{\varphi}_{N}=L\mu
^{-1}r^{-2} $. We neglected the precession of the orbital plane.
Accordingly, we have assumed that $(\mathbf{\hat{r}\times
\hat{v})\propto \hat{L}}$ for derivation of the angular equation in
Eq. (\ref{ang}). \footnote{The evolution of polar angle $\theta $
can be measured on the orbital plane, but this plane is not
conserved due to spin precession equation. Thus, the evolution of
$\theta $ can transform the inertial frame; see Ref. \cite{Wex}.} It
means that the inclination angle between the total angular momentum
$\mathbf{J}$ and the orbital angular momentum $\mathbf{L}$ is
constant because the evolution of the angle is squared in magnitude
of spin $d/dt(\mathbf{\hat{L}}\cdot \mathbf{\hat{J}})\approx
\mathcal{O}(S^{2})$ (for SSC II see Ref. \cite{recoil}). In other
words, it means that the orbital angular momentum $\mathbf{L}$ from
Eq. (\ref{momentum_exp}) determines the orbital plane instead of the
Newtonian angular momentum $\mathbf{L}_{\mathbf{N}}=\mu
(\mathbf{r\times v)}$ (for SSC II see Ref. \cite{Gopa}). If we
consider the unit angular momentum vector
$\mathbf{\hat{L}}=(\mathbf{\hat{r}\times \hat{v})(}1+\delta )$ from
Eq. (\ref{momentum_exp}), where $\delta $ is the leading-order
perturbation, then the angular equation is $\dot{\varphi}=$ $L\mu
^{-1}r^{-2}(1-\delta )$; see Eq. (\ref{ang}). We have assumed that
the scalar products $\mathbf{L}\cdot \mathbf{S}$ and
$\mathbf{L}\cdot {\mbox{\boldmath $\sigma$}}$ are constant because
they appear in the perturbative terms of $\mathcal{O}(c^{-2})$ or
the evolution of scalar products $\mathbf{S_{i}\cdot
L}_{\mathbf{N}}$ (or $\mathbf{S_{i}\cdot L}$) represents first-order
effects, see Ref. \cite{GPV3}. The radial equation agrees with the
expressions of Refs. \cite{Wex} and \cite{GPV3} for SSC I ($k=1$)
and SSC II ($k=2$), respectively. $\mathbf{L}$ and
$\mathbf{L}_{\mathbf{N}}$ appearing in perturbations are freely
interchangeable in scalar products because we have eliminated the
quadratic terms in spin $\mathcal{O}(S^{2})$.

Generally there are three angular equations with Euler angles (i.e.,
$\varphi ,\Upsilon ,\Theta _{N}$) for spin-orbit interaction given
by Ref. \cite{recoil} as
\begin{eqnarray} \dot{\varphi}
&=&\frac{L}{{\mu }r^{2}}\left( 1-\frac{\lambda
_{SO}}{2{L}^{2}}\right)
 -\cos \Theta _{N}\ \dot{\Upsilon},  \label{angle1} \\
\dot{\Upsilon} &=&\frac{\tan \varphi }{\sin \Theta _{N}}\dot{\Theta}_{N},
\label{angle2}
\end{eqnarray}
where $\lambda _{SO}$ is a shorthand notation for the SO
contributions in Eq. (\ref{angle1}) that we have computed for all
SSCs. $\lambda _{SO}=-2G\mu L/(c^{2}r)\left[
2\mathcal{S}\mathbf{+}3\left( 1-k\right) \Sigma \right]
+2(2k-1)EL/(c^{2}m)\Sigma $, which corresponds to the two SO
correction terms in Eq. (\ref{ang}). Here we used the original
notations of Ref. \cite{Wex}, where $\varphi =\psi$, $\Upsilon
=-\phi _{n}$, $\Theta _{N}=\mathbf{\hat{J}}\cdot
\mathbf{\hat{L}}_{\mathbf{N}}$ and $\Theta =\mathbf{\hat{J}}\cdot
\mathbf{\hat{L}}$ in the Hamiltonian formalism. It can be seen that
if we do not take the evolution of the angle $\Theta $ into account,
we get Eq. (\ref{ang}) from Eqs. (\ref{angle1}) and (\ref{angle2}).
In addition,
\begin{eqnarray}
\dot{\Theta}_{N} &=&-\frac{G\mu \left( \mathbf{S\cdot v}\right) }{c^{2}JL_{N}r^{3}}
\mathbf{[}4(\mathbf{S\cdot r)}+3({\mbox{\boldmath $\sigma$}\mathbf{\cdot r)}]}  \notag \\
&&{+\frac{3G\mu \dot{r}\left( {\mathbf{S\cdot r}}\right) }{c^{2}r^{4}}}\left[
{2(\mathbf{S\cdot r)}+(2-k)({\mbox{\boldmath
$\sigma$}}\mathbf{\cdot r)}}\right]  \notag \\
&&+\frac{G\mu \dot{r}\left( \mathbf{S}\cdot \mathbf{L}_{N}\right)
}{c^{2}JL_{N}^{2}r^{2}}[2(\mathbf{S\cdot
L}_{\mathbf{N}})+3(1-k)({\mbox{\boldmath $\sigma$}}\mathbf{\cdot
L}_{\mathbf{N}})]. \label{angle3}
\end{eqnarray}
It is important to know that the polar angle $\Theta $ does not
depend on SSC, but $\Theta _{N}$ does.
\begin{equation}
\dot{\Theta}=\frac{3G\mu \left( \nu ^{-1}-\nu \right)
}{2c^{2}Jr^{3}}\mathbf{L}\cdot \left( \mathbf{S}_{1}\times
\mathbf{S}_{2}\right) .  \label{angle3b}
\end{equation}
It can be seen that $\dot{\Theta}=0$ is relevant for two different
cases: (i) \textit{equal mass} ($\nu =1$) and (ii) \textit{single
spin} ($\mathbf{S}_{1}=0$ or $\mathbf{S}_{2}=0$) cases. The
evolutions of the angles of $\Theta $ and $\Theta _{N}$ are
quadratic in spin $\dot{\Theta},\dot{\Theta}_{N}\approx
\mathcal{O}(S^{2})$.

\section{Orbital motion}

Let us consider the radial motion which is characterized by Eq.
(\ref{radial}). We will use the generalized true anomaly $\chi$
parametrization \cite{GPV3},\cite{Kepler}
\begin{equation}
r=\frac{a_{r}(1-e_{r}^{2})}{1+e_{r}\cos \chi },  \label{param}
\end{equation}
where $a_{r}$ is the semimajor axis and $e_{r}$ is the radial
eccentricity. These parameters can be given by the turning points
$\dot{r}^{2}=0$. Thus we have found that the solution is in the form
$r_{{}_{{}_{min}^{max}}}=r_{\pm }(1+r_{\varepsilon })$, where
$r_{\pm }=-(Gm\mu \pm A)/(2E)=L^{2}/[\mu (Gm\mu \mp A)]$ is the
zeroth-order (or Newtonian) solution with $A^{2}=G^{2}m^{2}\mu
^{2}+2EL^{2}/\mu $ \footnote{There is a global minus misprint in
Ref. \cite{Kepler} for SSC II. The corrected equation is \textbf{\ }
$r_{{}_{{}_{min}^{max}}}=\frac{L^{2}}{\mu (Gm\mu \mp A)}+\frac{G\mu
(A\mp Gm\mu )\left( 4\mathbf{L} \cdot \mathbf{S}+3\mathbf{L}\cdot
{\mbox{\boldmath $\sigma$}}\right) }{2c^{2}L^{2}A}$.} and
$r_{\varepsilon }$ is the linear-order perturbation. Here $A$ is a
conserved quantity although it is not identical with the length of
the LRL vector which is only conserved for the Newtonian order.
Then, we get the turning points as
\begin{eqnarray}
r_{{}_{{}_{min}^{max}}} &=&\frac{L^{2}}{\mu (Gm\mu \mp A)}+(A\mp
Gm\mu )
\notag \\
&&\times \frac{4Gm\mu \mathcal{S}+\left[ 3Gm\mu \mp (1-2k)A\right]
\Sigma }{2c^{2}mLA}.
\end{eqnarray}
Here we used the notations for conserved scalar products
$\mathcal{S}=\mathbf{\hat{L}}\cdot \mathbf{S}$ and $\Sigma
=\mathbf{\hat{L}}\cdot {\mbox{\boldmath $\sigma$}}$ (the evolution
of these quantities are first-post Newtonian order effects, so we
could use them in linear-order terms; see Ref. \cite{GPV3}). The
relationship between the orbital elements and turning points is
$a_{r}=(r_{\max }+r_{\min })/2$ and $e_{r}=(r_{\max }-r_{\min
})/(r_{\max }+r_{\min })$, so the radial orbital parameters in all
SSCs are
\begin{eqnarray}
a_{r} &=&\frac{Gm\mu }{-2E}+\frac{G\mu }{c^{2}L}\left[ 2\mathcal{S}+(2-k)\Sigma \right] ,  \label{aa} \\
e_{r}^{2} &=&1+\frac{2EL^{2}}{G^{2}m^{2}\mu ^{3}}+\frac{4E}{c^{2}mL}\Biggl\{4
\left[ 1+\frac{EL^{2}}{G^{2}m^{2}\mu ^{3}}\right] \mathcal{S}  \notag \\
&&+\left[ 2\left( 2-k\right) +\frac{\left( 5-4k\right)
EL^{2}}{G^{2}m^{2}\mu ^{3}}\right] \Sigma \Biggr\}.  \label{ee}
\end{eqnarray}
Then, the conserved quantities with orbital elements are
\begin{eqnarray}
E &=&\frac{Gm\mu }{-2a_{r}}\Biggl\{1+\frac{G^{1/2}}{c^{2}m^{1/2}
a_{r}^{3/2}(1-e_{r}^{2})^{1/2}}  \notag \\
&&\times \left( 2\mathcal{S}+(2-k)\Sigma \right) \Biggr\}, \\
L^{2} &=&Gm\mu ^{2}a_{r}(1-e_{r}^{2})\Biggl\{1-\frac{G^{1/2}}{
c^{2}m^{1/2}a_{r}^{3/2}(1-e_{r}^{2})^{3/2}}  \notag \\
&&\times \left[
2(3+e_{r}^{2})\mathcal{S}+(5-k+3(1-k)e_{r}^{2})\Sigma \right]
\Biggr\}.
\end{eqnarray}
The time evolution of the generalized true anomaly from Eq.
(\ref{radial}) in terms of the orbital elements is
\begin{eqnarray}
\frac{dt}{d\chi } &=&\frac{r^{2}}{\sqrt{Gma_{r}(1-e_{r}^{2})}}
\Biggl\{1+\left( 2\mathcal{S}+(2-k)\Sigma \right)  \notag \\
&&\times \frac{G^{1/2}\left( e_{r}^{2}-3-2e_{r}\cos \chi \right)
}{2c^{2}m^{1/2}a_{r}^{3/2}(1-e_{r}^{2})^{3/2}}\Biggr\}.
\end{eqnarray}
After the integration we can get the result with \textit{eccentric
anomaly} $u$ parametrization, namely $r=a_{r}(1-e_{r}\cos u)$. In
other papers (e.g., \cite{Kepler}) it is indicated as $\xi$
\footnote{ Integration formulas for $e_{r}<1$ are $\int \frac{d\phi
}{(1+e_{r}\cos \chi )^{2}}=\frac{\left( u-e_{r}\sin u\right)
}{(1-e_{r}^{2})^{3/2}}$ and $\int \frac{\cos \phi d\phi
}{(1+e_{r}\cos \chi )^{2}}=-\frac{\left( e_{r}u-\sin u\right)
}{(1-e_{r}^{2})^{3/2}}$.}. Then we get the generalized Kepler
equation which contains the spin-orbit contributions in all SSCs as
\begin{equation}
n(t-t_{0})=u-e_{t}\sin u,  \label{Kepler}
\end{equation}
where we have introduced two orbital elements which are the
\textit{mean motion} $n$ and the \textit{time eccentricity} $e_{t}$
with conserved quantities ($E$, $L$, $\mathcal{S}$, and $\Sigma $),
\begin{eqnarray}
n &=&\frac{1}{Gm}\left( \frac{-2E}{\mu }\right) ^{3/2}, \\
e_{t}^{2} &=&1+\frac{2EL^{2}}{G^{2}m^{2}\mu ^{3}}+\frac{4E}{c^{2}mL}
\notag
\\
&&\times \Biggl\{2\mathcal{S}+\left[ 2-k+\frac{\left( 1-2k\right)
EL^{2}}{G^{2}m^{2}\mu ^{3}}\right] \Sigma \Biggr\}.
\end{eqnarray}
It can be seen that the mean motion does not contain SO terms and
the time eccentricity depends on SSC.

In the following let us consider the simple angular motion of the
binary systems which is described by Eq. (\ref{ang}). As we have
mentioned above, we solve the equation of motion in a noninertial
frame, which is the orbital plane. Thus, the angular equation from
Eq. (\ref{ang}) is
\begin{equation}
\dot{\varphi}=\frac{L}{\mu r^{2}}+\frac{\alpha }{r^{2}}+\frac{\beta
}{r^{3}}, \label{angleeq}
\end{equation}
where we have introduced the shorthand notations
\begin{eqnarray}
\alpha &=&{-}\frac{(2k-1)E\Sigma }{c^{2}m\mu }, \\
\beta &=&\frac{G\left[ 2\mathcal{S}+3(1-k)\Sigma \right] }{c^{2}}.
\end{eqnarray}
Using the generalized true anomaly parametrization in Eq.
(\ref{param}) the angular equation Eq. (\ref{angleeq}) can be
integrated in terms of the orbital elements
\begin{equation}
\frac{d\varphi }{d\chi }=1-\frac{G^{1/2}\left[ 4\mathcal{S}+3\Sigma
-\left( 1-2k\right) \Sigma e_{r}\cos \chi \right]
}{c^{2}m^{1/2}a_{r}^{3/2}(1-e_{r}^{2})^{3/2}}\ .
\end{equation}
After the integration we get the angular motion as (see Ref.
\cite{MFV} for the first post-Newtonian corrections)
\begin{equation}
\varphi -\varphi _{0}=K\chi -Q\sin \chi ,  \label{GPVparam}
\end{equation}
where $\varphi _{0}$ is the integration constant. We have also
introduced some shorthand notations with conserved quantities
\begin{eqnarray}
K &=&1-\frac{G^{2}m\mu ^{3}\left( 4\mathcal{S}+3\Sigma \right)
}{c^{2}L^{3}},
\\
Q &=&\frac{G\mu ^{2}\left( 2k-1\right) A\Sigma }{c^{2}L^{3}}.
\end{eqnarray}

There is another solution for the angular evolution in literature,
which is introduced by Damour and Deruelle in Ref. \cite{DD85} using
the \textit{conchoidal transformation.} In this parametrization
there is a third eccentricity $e_{\theta }$. If we use the
conchoidal transformation $r=\tilde{r}$ $+\beta /(2\tilde{L})$ with
$\tilde{L}=L/\mu +\alpha $ in Eq.(\ref{angleeq}), then the angular
equation has the simple form (like the Newtonian equation for the
angular motion)
\begin{equation}
\dot{\varphi}=\frac{\tilde{L}}{\tilde{r}^{2}}.  \label{angleDD}
\end{equation}
The integration of this angular equation with the generalized
eccentric anomaly parametrization $r=a_{r}(1-e_{r}\cos u)$, where we
used the \textit{deformed} parametrization
\begin{equation}
\tilde{r}=\tilde{a}\left( 1-\tilde{e}\cos u\right) ,
\label{deform_concoidal}
\end{equation}
with $\tilde{a}=a_{r}-\beta /(2\tilde{L})$ and
$\tilde{e}=a_{r}e_{r}/\tilde{a}$ as shorthand notations is
straightforward. With the help of Eqs. (\ref{deform_concoidal}) and
(\ref{Kepler}) in Eq. (\ref{angleDD}) we get
\begin{equation}
\frac{d\varphi }{du}=\frac{\tilde{L}}{n\tilde{a}^{2}(1-e_{\theta
}\cos u)},
\end{equation}
where we have introduced the \textit{angular eccentricity}
$e_{\theta }=2\tilde{e}-e_{t}$ as an orbital parameter. After the
integration we get
\begin{equation}
\varphi -\varphi _{0}=(1+\tilde{k})v,  \label{DDparam}
\end{equation}
where $\tilde{k}=\tilde{L}/(n\tilde{a}^{2}\sqrt{1-e_{\theta
}^{2}})-1$ is the \textit{pericenter drift} and $v=2\arctan \left(
\sqrt{(1+e_{\theta })/(1-e_{\theta })}\tan u/2\right) $ is a similar
Damour-Deruelle true anomaly. Finally, we add the angular orbital
elements with conserved quantities, as
\begin{eqnarray}
\tilde{k} &=&{-}\frac{G^{2}m\mu ^{3}\left( 4\mathcal{S}+3\Sigma \right) }{c^{2}L^{3}},  \label{k} \\
e_{\theta }^{2} &=&1+\frac{2EL^{2}}{G^{2}m^{2}\mu ^{3}}  \notag \\
&&+\left( 1+\frac{EL^{2}}{G^{2}m^{2}\mu ^{3}}\right)
\frac{4E(4\mathcal{S}+3\Sigma {)}}{c^{2}mL}.  \label{e_theta}
\end{eqnarray}
\qquad

The Damour-Deruelle angular orbital parameters $e_{\theta }$ and
$\tilde{k}$ in Eqs. (\ref{k}) and (\ref{e_theta}) are not
SSC-dependent. This angular motion does not agree with cases of SSC
I/II in Refs. \cite{Wex}, \cite{CK}, and \cite{KJ} because in these
cases they only considered the Newtonian term the first term in Eq.
(\ref{ang}), but we have the same angular motion in Ref.
\cite{Gopa}, which is identical with the paper of Ref.
\cite{Tessmer} for the eccentric case (see the Appendix C).

Both parametrizations are equivalent to each other. Apparently Eq.
(\ref{GPVparam}) depends on SSC, but if we use the eccentric anomaly
parametrization $u$, the SSC dependence disappears. The
relationships between quantities for the angular motion are
\begin{eqnarray}
K &=&1+\tilde{k} \\
Q &=&\frac{Gm\mu ^{2}A\rho }{2EL^{2}},
\end{eqnarray}
where we have introduced $\rho =1-e_{r}/e_{\theta }$ as a shorthand
notation.

\section{Dissipation under gravitational radiation}

The energy and the orbital angular momentum change due to the
gravitational radiation at 2.5 PN order. The instantaneous losses
for the spin-orbit interaction were given by Kidder \cite{Kidder}
using SSC I. Some authors calculated the averaged losses for SSC
I/II \cite{RS,GPV3,CK}. We compute these averaged losses for all
SSCs including the missing SSC III. The multipolar momenta are
necessary for computation of the energy and the angular momentum
losses up to the SO order. The mass $\mathcal{I}_{ij}$ and current
$\mathcal{J}_{ij}$ quadrupole momentums in relative Descartes
coordinates are
\begin{eqnarray}
\mathcal{I}_{ij} &=&\mu \left( r_{i}r_{j}\right) ^{STF}  \notag \\
&&+\frac{2\eta }{3c^{2}}\left( \varepsilon _{ipq}\left[
(1+3k)r_{j}v_{p}-2v_{j}r_{p}\right] {{\mbox{$\sigma$}}}_{q}\right)
^{STF},
\label{mass_momenta} \\
\mathcal{J}_{ij} &=&-\eta \delta m\left( \varepsilon
_{ipq}r_{j}r_{p}v_{q}\right) ^{STF}  \notag \\
&&+\frac{3\mu }{2\delta m}\left( r_{i}\left[
S_{j}-{{\mbox{$\sigma$}}}_{j} \right] \right) ^{STF},
\label{current_momenta}
\end{eqnarray}
where $\eta =\mu /m$ is the symmetric mass ratio, $x_{i}$ are
relative coordinates, $v_{i}=\dot{x}_{i}$ is the relative velocity
of the binary, $S_{i}$ and ${{\mbox{$\sigma$}}}_{i}$ are the
coordinates of the spin vector $\mathbf{S}$ and
$\mathbf{\mbox{\boldmath $\sigma$}}$, respectively, the mass
difference $\delta m=m_{1}-m_{2}$ (choosing $m_{1}\geqq m_{2}$ by
convention) and $\varepsilon _{ijk}$ is the Levi-Civita symbol. The
last term is apparently singular for equal masses in Eq.
(\ref{current_momenta}) because it can be expressed as
$(3/2)Gm(r_{i}\eta _{j})^{STF}$ with another spin vector
${\mbox{\boldmath $\eta$}}=\mu
(\mathbf{S}_{1}/m_{1}-\mathbf{S}_{2}/m_{2})/(Gm^{2})$ (see Table I).
It can be proved that the current angular momentum
$\mathcal{J}_{ij}$ does not depend on SSC. Here $STF$ means the
indices of the momentums $\mathcal{I}_{ij}$ and $\mathcal{J}_{ij}$
are symmetric-trace-free. Thus, the instantaneous energy and the
angular momentum losses up to the SO order are given by \cite{KWW}
\begin{eqnarray}
\frac{dE}{dt} &=&-\frac{G}{5c^{5}}\left( \mathcal{\dddot{I}}_{ij}\mathcal
{\dddot{I}}_{ij}+\frac{16}{9c^{2}}\mathcal{\dddot{J}}_{ij}\mathcal{\dddot{J}}_{ij}\right) , \\
\frac{dL}{dt} &=&-\frac{2G}{5c^{5}}\varepsilon _{ipq}\left(
\mathcal{\ddot{J}}_{pj}\mathcal{\dddot{I}}_{qj}+\frac{16}{9c^{2}}\mathcal{\ddot{J}}_{pj}\mathcal{\dddot{J}}_{qj}\right)
\hat{L}_{i},
\end{eqnarray}
where repeated indices indicate summation, dots over multipolar
moments mean time derivatives, and $\hat{L}_{i}$ denotes the
components of the unit angular momentum vector in Eq.
(\ref{momentum_exp}). Then we get the instantaneous losses for
different SSCs
\begin{eqnarray}
\frac{dE}{dt} &=&\frac{8G^{3}m^{2}\mu ^{2}}{15c^{5}r^{4}}\left( 11\dot{r}^{2}-12v^{2}\right)  \notag \\
&&-\frac{8G^{3}m\mu L}{15c^{7}r^{6}}\Biggl\{\left[ 27\dot{r}^{2}-37v^{2}-12\frac{Gm}{r}\right] \mathcal{S}  \notag \\
&&+\biggl [3(22k-5)\dot{r}^{2}-(48k-5)v^{2}  \notag \\
&&+4(6k-5)\frac{Gm}{r}\biggr ]\Sigma \Biggr\},
\label{instdE} \\
\frac{dL}{dt} &=&\frac{8G^{2}m\mu L}{5c^{5}r^{5}}\left( 3\dot{r}^{2}-2v^{2}-\frac{2Gm}{r}\right)  \notag \\
&&+\frac{12G^{2}\mu ^{2}}{45c^{7}r^{7}}\Biggl\{\biggl [6\left( 3\dot{r}^{2}v^{2}-4\dot{r}^{4}+v^{4}\right)  \notag \\
&&-26\frac{Gm}{r}\left( \dot{r}^{2}-v^{2}\right) -6\frac{G^{2}m^{2}}{r^{2}}\biggr ]\mathcal{S}  \notag \\
&&+\biggl [6(16-21k)\dot{r}^{2}v^{2}-(78-90k)\dot{r}^{4}  \notag \\
&&-(17-36k)v^{4}+\frac{Gm}{r}[(7-24k)\dot{r}^{2}  \notag \\
&&-8(1-3k)v^{2}]-5\frac{G^{2}m^{2}}{r^{2}}\biggr ]\Sigma \Biggr\},
\label{instdL}
\end{eqnarray}
where $\mathcal{S=}\mathbf{\hat{L}}\cdot \mathbf{S}$ and $\Sigma
\mathcal{=}\mathbf{\hat{L}}\cdot {\mbox{\boldmath $\sigma$}}$. It
can be seen that our results are equivalent with that of Ref.
\cite{Kidder} for $k=1$ and Ref. \cite{RS} for $k=1/2$.

\begin{figure}[!hb]
\includegraphics[width=0.9\columnwidth]{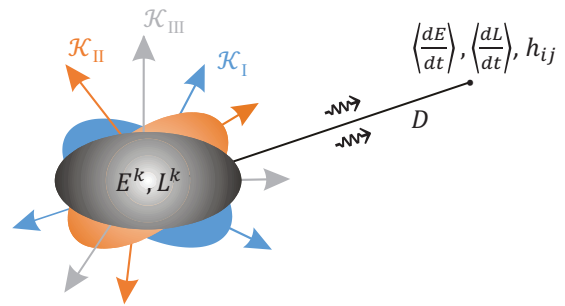}
\caption{Dissipative quantities in the different SSCs. The compact
binary system is characterized by the energy $E$ and the magnitude
of the orbital angular momentum $L$, which depend on SSC (here $k$
means the SSC dependence). Here $D$ is the luminosity distance and
$\mathcal{K}_{I,II,III}$ are the different frames for SSCs. Thus,
the averaged energy and the orbital angular momentum losses due to
gravitational radiation depend on SSC, however the leading-order
spin-orbit waveform does not depend on SSC.} \label{fig3}
\end{figure}

The instantaneous energy and the angular momentum losses depend on
SSC in Eqs. (\ref{instdE}) and (\ref{instdL}), so these formulas
involve parameter $k$. If we use the $E$ and $L$ conserved
quantities instead of $r$, $v$ and $\dot{r}$, the dependence on
parameter $k$ remains. On the other hand, if we average these
formulas for one Newtonian orbital period (see Ref. \cite{BOC70}),
the explicit $k$ dependence disappears, but $E$ and $L$ depend on
SSC as in Eqs. (\ref{energy_exp}) and (\ref{momentum_exp})(see Fig.
\ref{fig3}).
\begin{eqnarray}
\left\langle {\frac{dE}{dt}}\right\rangle &=&-\frac{G^{2}m(-2E\mu )^{3/2}}{15c^{5}L^{7}}(148E^{2}L^{4}+732G^{2}m^{2}\mu ^{3}EL  \notag \\
&&+425G^{4}m^{4}\mu ^{6})+\frac{G^{2}(-2E\mu )^{3/2}}{10c^{7}L^{9}}\biggl [(520E^{3}L^{6}  \notag \\
&&+10740G^{2}m^{2}\mu ^{3}E^{2}L^{4}+24990G^{4}m^{4}\mu ^{6}EL^{2}  \notag \\
&&+12579G^{6}m^{6}\mu ^{9})\mathcal{S}+(256E^{3}L^{6}  \notag \\
&&+6660G^{2}m^{2}\mu ^{3}E^{2}L^{4}+16660G^{4}m^{4}\mu ^{6}EL^{2}  \notag \\
&&+8673G^{6}m^{6}\mu ^{9})\Sigma \biggr ],  \label{avealossE} \\
\left\langle {\frac{dL}{dt}}\right\rangle &=&-\frac{4G^{2}m(-2E\mu )^{3/2}}{5c^{5}L^{4}}(14EL^{2}+15G^{2}m^{2}\mu ^{3})  \notag \\
&&+\frac{G^{2}(-2E\mu )^{3/2}}{15c^{7}L^{6}}\biggl [(1188E^{2}L^{4}
\notag
\\
&&+6756G^{2}m^{2}\mu ^{3}EL^{2}+5345G^{4}m^{4}\mu ^{6})\mathcal{S}  \notag \\
&&+(772E^{2}L^{4}+4476G^{2}m^{2}\mu ^{3}EL^{2}  \notag \\
&&+3665G^{4}m^{4}\mu ^{6})\Sigma \biggr ].  \label{avelossL}
\end{eqnarray}
We will compute the SO contributions of the waveform for different
SSCs in the next chapter.

\section{Waveform}

We need the current octopole momentum $\mathcal{J}_{ijk}$ for the
computation of the waveform $h_{ij}$ (see Eq. (3.20b) in Ref.
\cite{Kidder}). Thus, the $\mathcal{J}_{ijk}$ does not depend on SSC
up to the SO-order as
\begin{eqnarray}
\mathcal{J}_{ijk} &=&\mu (1-3\eta )\left( r_{i}r_{j}\varepsilon
_{kpq}r_{p}v_{q}\right) ^{STF}  \notag \\
&&+2\eta \left( r_{i}r_{j}{{\mbox{$\sigma$}}}_{k}\right) ^{STF}.
\label{octo}
\end{eqnarray}
The second term in Eq. (\ref{octo}) is relevant for the computation
of the waveform, as the first term only appears in the next PN-order
corrections. The waveform up to the SO order is computed by
\begin{eqnarray}
h_{ij} &=&\frac{2G}{c^{4}D}\biggl [\mathcal{\ddot{I}}_{ij}+\frac{4}{3c^{2}}
\varepsilon _{kl(i}\mathcal{\ddot{J}}_{j)k}N_{l}  \notag \\
&&+\frac{1}{2c^{2}}\varepsilon
_{kl(i}\mathcal{\dddot{J}}_{j)km}N_{l}N_{m}\biggr ]_{TT},
\end{eqnarray}
where $D$ is the distance between the source and observer, $N_{k}$
are the components of the unit vector $\mathbf{N}$ which points from
the source to the observer, and $(..)_{TT}$ means the
\textit{transverse-traceless} transformation (we have omitted the
pure relativistic PN corrections, so the first terms in Eqs.
(\ref{octo},\ref{current_momenta}) can be neglected).

The gravitational waveforms for all SSCs (here we have neglected the
pure relativistic corrections $P^{0.5-1.5}Q_{ij}$ which are given in
\cite{WW},\cite{Wiseman}) are given as
\begin{equation}
h_{ij}=\frac{2G\mu }{c^{4}D}\left[
Q_{ij}+PQ_{ij}^{SO}+P^{1.5}Q_{ij}^{SO}\right] _{TT},
\label{waveform}
\end{equation}
with
\begin{eqnarray}
Q_{ij} &=&2\left( v_{i}v_{j}-\frac{Gm}{r^{3}}r_{i}r_{j}\right) ,
\label{Leading} \\
PQ_{ij}^{SO} &=&\frac{2m}{r^{3}\delta m}\Bigl\{\left[ \left(
{\mbox{\boldmath $\sigma$}-\mathbf{S}}\right) {\times
}\mathbf{N}\right]
_{(i}r_{j)}\Bigr\},  \label{LeadingSO} \\
P^{1.5}Q_{ij}^{SO}
&=&\frac{2}{r^{3}}\Biggl\{\frac{3r_{i}r_{j}}{r^{2}}(\mathbf{r\times
v})\mathbf{\cdot }\left[ 2\mathbf{S}+(1+k)\mbox{\boldmath
$\sigma$}\right]   \notag \\
&&-r_{(i}[\mathbf{v\times (}4\mathbf{S}+(3+2k)\mbox{\boldmath
$\sigma$})]_{j)}  \notag \\
&&-2kv_{(i}(\mathbf{r\times }\mbox{\boldmath
$\sigma$})_{j)}+\frac{6\dot{r}}{r}r_{(i}[\mathbf{r\times
(S}+\mbox{\boldmath
$\sigma$})]_{j)}  \notag \\
&&+\biggl [\left( \frac{3\dot{r}}{r}\mathbf{r}-2\mathbf{v}\right) (\mathbf{N\cdot r})  \notag \\
&&-2\mathbf{r}(\mathbf{N\cdot v})\biggr ]_{(i}(\mbox{\boldmath
$\sigma$}\mathbf{\times N)}_{j)}\Biggr\},  \label{nextLeadingSO}
\end{eqnarray}
where we have used the following formulas which are valid for any
TT-tensor and $\mathbf{a}$ and $\mathbf{b}$ vectors
\begin{eqnarray}
\left( \delta _{ij}\right) _{TT} &=&\left( N_{i}b_{j}\right)
_{TT}=0,  \notag
\\
\left[ b_{i}\left( \mathbf{a\times N}\right) _{j}\right] _{TT}
&=&\left[ a_{i}\left( \mathbf{b\times N}\right) _{j}\right] _{TT},
\end{eqnarray}
where $\delta _{ij}$ is the Kronecker delta function. The tensor
$Q_{ij}$ is the zeroth-order waveform, the $PQ_{ij}^{SO}$ is the
leading-order SO contribution (which does not contain terms
$\mathcal{O}(\mathbf{N})$) and the $P^{1.5}Q_{ij}^{SO}$ is the
next-to-leading-order SO contribution (which is proportional to
terms $\mathcal{O}(\mathbf{N}^{0})$ and
$\mathcal{O}(\mathbf{N}^{2})$ ) to the waveform. The leading-order
SO contribution $PQ_{ij}^{SO}$ is singular for equal-masses, since
$({\mbox{\boldmath $\sigma$}-\mathbf{S)/}}\delta m{\mathbf{\equiv
S}}_{2}/m_{2}-{\mathbf{S}}_{1}/m_{1}$. Here we can use the
spinvector $\mathbf{\Delta}$ of Kidder (see Table I). It is
transparent that for $k=1$ we retain the SSC I case as in the
classical paper of \cite{Kidder}.

\begin{widetext}

\begin{table}[h!]

\resizebox{\textwidth}{!}{
\begin{tabular}{ccccc}
 \hline\hline
spinvectors & $(\mathbf{S}_{1},\mathbf{S}_{2})$ &
$({\mathbf{S,}\mbox{\boldmath
$\sigma$})}$ & $({\mathbf{S,}}\mathbf{\Delta )}$ & $(\boldsymbol{\chi }_{s},\boldsymbol{\chi }_{a})$ \\
$(\mathbf{S}_{1},\mathbf{S}_{2})$ & - & $
\begin{array}{c}
{\mathbf{S=S}}_{1}+{\mathbf{S}}_{2}, \\
{\mbox{\boldmath
$\sigma$}}=\frac{m_{2}}{m_{1}}\mathbf{S}_{1}+\frac{m_{1}}{m_{2}}\mathbf{S}_{2}
\end{array}
$ & $
\begin{array}{c}
{\mathbf{S=S}}_{1}+{\mathbf{S}}_{2}\text{, } \\
\mathbf{\Delta }=m\left(
\frac{\mathbf{S}_{2}}{m_{2}}-\frac{\mathbf{S}_{1}}{m_{1}}\right)
\end{array}
$ & $
\begin{array}{c}
\boldsymbol{\chi }_{s}=\frac{1}{2}\left( \frac{\mathbf{S}_{2}}{m_{2}^{2}}+\frac{\mathbf{S}_{1}}{m_{1}^{2}}\right) \text{, } \\
\boldsymbol{\chi }_{a}=\frac{1}{2}\left(
\frac{\mathbf{S}_{2}}{m_{2}^{2}}-\frac{\mathbf{S}_{1}}{m_{1}^{2}}\right)
\end{array}
$ \\ \hline $({\mathbf{S,}\mbox{\boldmath $\sigma$})}$ & $
\begin{array}{c}
{\mathbf{S}}_{1}{\mathbf{=}}\frac{\nu
^{-1}{\mathbf{S}}-{\mbox{\boldmath
$\sigma$}}}{\nu ^{-1}-\nu }, \\
{\mathbf{S}}_{2}=\frac{\nu {\mathbf{S}}-{\mbox{\boldmath
$\sigma$}}}{\nu -\nu ^{-1}}
\end{array}
$ & - & $\mathbf{\Delta =}\frac{1+\nu }{1-\nu }\left(
{\mbox{\boldmath $\sigma$}-\mathbf{S}}\right) $ & $
\begin{array}{c}
\boldsymbol{\chi }_{s}\mathbf{=}\frac{{{(1+\nu )}^{2}\mathbf{S+}(1-\nu ^{2})\mathbf{\Delta }}}{2m^{2}{\nu }}\text{, } \\
\boldsymbol{\chi }_{a}\mathbf{=}\frac{\left( 1+\nu \right)
^{2}\left( \mathbf{S-\Delta }\right) }{2m^{2}{\nu }}\end{array}
$ \\
$({\mathbf{S,}}\mathbf{\Delta )}$ & $
\begin{array}{c}
{\mathbf{S}}_{1}{\mathbf{=}}\frac{\left( \nu ^{-1}-\nu \right)
{\mathbf{S}}\mathbf{-}\left( 1-\nu \right) \mathbf{\Delta }}{\nu
^{-1}+1-\nu -\nu ^{2}},
\\
{\mathbf{S}}_{2}=\frac{\left( 1-\nu ^{2}\right) {\mathbf{S+}}\left(
1-\nu \right) \mathbf{\Delta }}{\nu ^{-1}+1-\nu -\nu
^{2}}\end{array} $ & ${\mbox{\boldmath
$\sigma$}}\mathbf{=S+}\frac{1-\nu }{1+\nu }\mathbf{\Delta }$ & - &
$\begin{array}{c} \boldsymbol{\chi }_{s}\mathbf{=}\frac{{(1+\nu
)}^{2}{\mbox{\boldmath
$\sigma$}}}{2m^{2}{\nu }}\text{, } \\
\boldsymbol{\chi }_{a}\mathbf{=}\frac{2\left( 1+\nu \right)
{\mathbf{S-(}}1+\nu ^{2})\left( 1+\nu \right) {\mbox{\boldmath
$\sigma$}}}{2m^{2}{\nu }\left( 1-\nu \right) }\end{array}$ \\
$(\boldsymbol{\chi }_{s},\boldsymbol{\chi }_{a})$ &
$\begin{array}
{c} \mathbf{S}_{1}=\frac{\nu m^{2}\left(
\boldsymbol{\chi }_{s}-\boldsymbol{\chi
}_{a}\right) }{(1+\nu )^{2}}\text{, } \\
\mathbf{S}_{2}=\frac{m^{2}\left( \boldsymbol{\chi
}_{s}+\boldsymbol{\chi }_{a}\right) }{(1+\nu )^{2}}\end{array} $ & $
\begin{array}{c}
{\mathbf{S}}\mathbf{=}\frac{m^{2}\left[ \left( 1-\nu ^{2}\right)
\boldsymbol{\chi }_{a}+{\mathbf{(}}1+\nu ^{2})\boldsymbol{\chi
}_{s}\right] }{(1+\nu
)^{2}}, \\
{\mbox{\boldmath $\sigma$}\mathbf{=}}\frac{2\nu
m^{2}\boldsymbol{\chi }_{s}}{(1+\nu )^{2}}\end{array} $ & $
\begin{array}{c}
{\mathbf{S=}}\frac{\nu m^{2}\left[ \left( \nu ^{-1}-\nu \right)
\boldsymbol{\chi }_{a}+{\mathbf{(}}\nu ^{-1}+\nu )\boldsymbol{\chi
}_{s}\right] }{(1+\nu
)^{2}}\text{, } \\
\mathbf{\Delta }=\frac{m^{2}\left[ \left( 1-\nu ^{2}\right)
\boldsymbol{\chi }_{a}+{\mathbf{(}}1+\nu ^{2})\boldsymbol{\chi
}_{s}\right] }{1-\nu ^{2}}\end{array} $ & - \\
\hline\hline\label{tab}
\end{tabular}
}

\caption{Different notations for spinvectors. One of the most widely
used notations are the total ${\mathbf{S}}$ and the weighted
${\mbox{\boldmath $\sigma$}}$ spins in \protect\cite{GPV3}.
${\mathbf{S\equiv } \mbox{\boldmath $\zeta$}}$ and ${\mbox{\boldmath
$\sigma$}\mathbf{\equiv } \mbox{\boldmath $\xi$}}$ in
\protect\cite{KWW} \protect\cite{Kidder}, and \protect\cite{RS} (but
here a factor $Gm^{2}$ is used in definitions, thus
${\mbox{\boldmath
$\zeta$}}=({\mathbf{S}}_{1}+{\mathbf{S}}_{2})/Gm^{2}$ and
${\mbox{\boldmath $\xi$}}=\left( \protect\nu
\mathbf{S}_{1}+\protect\nu ^{-1}\mathbf{S}_{2}\right) /Gm^{2}$ where
$\nu=m_{2}/m_{1}$ is the mass ratio parameter). The total and other
combinations of weighted spins $\mathbf{\Delta =}(m/\protect\delta
m)\mathbf{(}\mbox{\boldmath $\sigma$}-\mathbf{S)}$ are also used in
\protect\cite{Kidder},\protect\cite{Blanchet}, $\mathbf{\Sigma
\equiv \Delta }$.\ Some authors used an effective spin combination
$\mathbf{S}_{eff}=2{\mathbf{S+(}}3/2){\mbox{\boldmath $\sigma$}}$,
which is a convenient notation for SSC II in
\cite{Wex,Gopa,Tessmer,Racine,CK}. There are other symmetrized spin
quantities ${\mbox{\boldmath
$\chi_s$}}=({\mathbf{S}}_{2}/m_{2}^{2}+{\mathbf{S}}_{1}/m_{1}^{2})/2$
and ${\mbox{\boldmath
$\chi_a$}}=({\mathbf{S}}_{2}/m_{2}^{2}-{\mathbf{S}}_{1}/m_{1}^{2})/2$
which are useful notations for the dimensionless angular-momentum
\textit{Kerr parameters} of the individual bodies in Refs.
\protect\cite{WW},\protect\cite{Tagoshi}. }

\end{table}

\end{widetext}

\section{Summary}

We presented the spin supplementary conditions for the leading-order
spin-orbit contribution of compact binaries. The Lagrangian contains
acceleration dependent terms in some cases of SSC. Thus we have to
use the Ostrogradsky dynamics for generalized Lagrangian. We have
shown some procedures of the elimination of the acceleration from
the Lagrangian, i.e. the method of the double zero and constrained
dynamics in Appendices. We constructed the generalized Hamiltonian
function with the presence of high-order canonical moments and
computed the generalized Hamilton's equations.

Our radial and angular motion of the compact binaries represent the
SSC dependence of any orbital parameters for eccentric orbits. We
calculated the energy and the orbital angular momentum losses due to
gravitational radiation in each SSC, and we concluded that the
dependence of SSC apparently disappears, since we use averaging over
one orbital period. \textit{However, these expressions are
SSC-dependent because the energy and the orbital angular momentum
depend on SSC, see Eqs. (\ref{energy}) and (\ref{momentum})}.

Nevertheless, we calculated the leading-order gravitational waveform
contains the spin-orbit corrections. \textit{It has been proven that
the leading-order spin-orbit does not depend on SSC but the
next-to-leading-order spin-orbit contribution does.}

\section{Acknowledgements}

I would like to thank P\'{e}ter Forg\'{a}cs and M\'{a}ty\'{a}s
Vas\'{u}th for some critical reading of the manuscript. This work
was supported by the Postdoctoral Fellowship Programme of the
Hungarian Academy of Sciences and Hungarian Scientific Research Fund
(OTKA) Grant No. 116892.

\appendix

\subsection{Appendix A: The elimination of acceleration: \textit{Constrained dynamics}}

Constrained dynamics arose from the degenerate Lagrangian developed
by Dirac, Anderson and Bergmann (see \cite{Dirac,AB}). The simple
acceleration-dependent Lagrangian of a relativistic spinning body
studied by \cite{Riewe} and \cite{Ellis} leads to constrained
dynamics. The Dirac formalism for constrained Hamiltonian of a
spherical spinning top interacting with Poisson brackets was given
by Ref. \cite{HRT}.

We introduce two new variables using the method of Lagrange
multipliers where ${\mbox{\boldmath $\lambda$}}\mathbf{=\dot{r}}$
for the acceleration term, and ${\mbox{\boldmath $\delta$}}$ is a
multiplier in the Lagrangian. The transformation of the Lagrangian
is $\mathcal{L}(\mathbf{r},\mathbf{v,a} )\rightarrow
\mathcal{L}^{\ast }(\mathbf{r},\mathbf{v,{\mbox{\boldmath
$\lambda$}},{\mbox{\boldmath ${\dot{\lambda}}$}},}{\mbox{\boldmath
$\delta$}})$ as
\begin{eqnarray}
\mathcal{L}^{\ast } &=&\frac{\mu }{2}\mathbf{v}^{2}+\frac{Gm\mu
}{r}+\frac{G\mu }{c^{2}r^{3}}
\mathbf{v\cdot }\left[ \mathbf{r}\times \left( 2\mathbf{S}+(1+k){\mbox{\boldmath $\sigma$}}\right) \right]   \notag \\
&&+\frac{(2k-1)\mu }{2c^{2}m}\mathbf{v}\cdot ({\mbox{\boldmath
${\dot{\lambda}}$}}\times {\mbox{\boldmath
$\sigma$})}+{\mbox{\boldmath $\delta$}}\mathbf{\cdot
}(\mathbf{v}-{\mbox{\boldmath $\lambda$}}).
\end{eqnarray}
Then, the Euler-Lagrange equations are
\begin{eqnarray}
\mu \mathbf{a} &=&-\frac{Gm\mu }{r^{3}}\mathbf{r}-\frac{2G\mu
}{c^{2}r^{3}}\mathbf{v}\times \left(
2\mathbf{S}+(1+k){\mbox{\boldmath $\sigma$}}\right)
\notag \\
&&+\frac{3G\mu }{c^{2}r^{5}}\mathbf{r}\left[ \left( \mathbf{r}\times
\mathbf{v}\right)
\mathbf{\cdot }\left( 2\mathbf{S}+(1+k){\mbox{\boldmath $\sigma$}}\right) \right]   \notag \\
&&\mathbf{+}\frac{3G\mu \dot{r}}{c^{2}r^{4}}\mathbf{r}\times \left( 2\mathbf{S}+(1+k){\mbox{\boldmath $\sigma$}}\right)   \notag \\
&&-\frac{(2k-1)\mu }{2c^{2}m}({\mbox{\boldmath
${\ddot{\lambda}}$}}\times {\mbox{\boldmath
$\sigma$})-\mbox{\boldmath ${\dot{\delta}}$}}\mathbf{,}  \notag \\
0 &=&{\mbox{\boldmath $\delta$}}-\frac{(2k-1)\mu
}{2c^{2}m}(\mathbf{a}\times
{\mbox{\boldmath $\sigma$}),}  \notag \\
0 &=&\mathbf{v}-{\mbox{\boldmath $\lambda$}}\mathbf{.}\label{ELcon}
\end{eqnarray}
Using these equations, we have derived the acceleration of Eq.
(\ref{acc}). It can be seen that the Lagrangian is degenerate, so we
have to construct the constrained dynamics for this case.We compute
the conjugate momenta as
\begin{equation}
\mathbf{p}_{\mathbf{r}}=\frac{\partial \mathcal{L}^{\ast }}{\partial
\mathbf{\dot{r}}},\qquad \mathbf{p}_{\bm{\lambda }}=\frac{\partial
\mathcal{L}^{\ast }}{\partial {\mbox{\boldmath
${\dot{\lambda}}$}}},\qquad \mathbf{p}_{\bm{\delta }}=\frac{\partial
\mathcal{L}^{\ast }}{\partial {\mbox{\boldmath ${\dot{\delta}}$}}},
\end{equation}
then
\begin{eqnarray}
\mathbf{p}_{\mathbf{r}} &=&\mu \mathbf{v+}\frac{G\mu
}{c^{2}r^{3}}\mathbf{r}
\times \left( 2\mathbf{S}+(1+k){\mbox{\boldmath $\sigma$}}\right)   \notag \\
&&+\frac{(2k-1)\mu }{2c^{2}m}{\mbox{\boldmath
${\dot{\lambda}}$}}\times {\mbox{\boldmath
$\sigma$}}+{\mbox{\boldmath $\delta$}}\mathbf{,}  \notag \\
\mathbf{p}_{\bm{\lambda }} &=&-\frac{(2k-1)\mu
}{2c^{2}m}\mathbf{v}\times {\mbox{\boldmath $\sigma$},}
\end{eqnarray}
and the first kind of subsidiary condition is
\begin{equation}
\bm{\phi}_{1}\doteq \mathbf{p}_{\bm{\delta }}\approx 0,
\end{equation}
where the symbol $\approx $ denotes the weak equality (see Ref.
\cite{Dirac}). The second kind of condition is
\begin{eqnarray}
\bm{\phi}_{2}\doteq &\mathbf{v}-{\mbox{\boldmath $\lambda$}}  \notag \\
&\mathbf{=}&\frac{\mathbf{p}_{\mathbf{r}}}{\mu
}-\frac{G}{c^{2}r^{3}}\mathbf{r}\times \left(
2\mathbf{S}+(1+k){\mbox{\boldmath $\sigma$}}\right) \notag
\\
&&-\frac{(2k-1)}{2c^{2}m}{\mbox{\boldmath ${\dot{\lambda}}$}}\times
{\mbox{\boldmath $\sigma$}}-\frac{{\mbox{\boldmath $\delta$}}}{\mu
}-{\mbox{\boldmath $\lambda$}}\approx 0.
\end{eqnarray}
A new further condition can be given as
\begin{equation}
\bm{\phi}_{3}\doteq \bm{\dot\phi}_{2}\approx 0.
\end{equation}
Then, the Hamiltonian is
\begin{equation}
\mathcal{H}=\mathcal{H}_{0}+\underset{i=1}{\overset{3}{\sum
}}\mathbf{c}_{i}\cdot {\bm{\phi}_{i}},
\end{equation}
where $\mathbf{c}_{i}$ are arbitrary multipliers and
\begin{equation}
\mathcal{H}_{0}=\mathbf{p}_{\mathbf{r}}\cdot
\mathbf{\dot{r}}+\mathbf{p}_{\bm{\lambda }}\cdot {\mbox{\boldmath
${\dot{\lambda}}$}}-\mathcal{L}^{\ast }.
\end{equation}
It can be seen that, the final Hamiltonian is
\begin{eqnarray}
\mathcal{H} &=&\frac{\mathbf{p}_{\mathbf{r}}^{2}}{2\mu }-\frac{Gm\mu
}{r}+\frac{G(2k-1)}{2c^{2}r^{3}}\mathbf{p}_{\mathbf{r}}\cdot
(\mathbf{r}\times {\mbox{\boldmath
$\sigma$})}  \notag \\
&&-\frac{G}{c^{2}r^{3}}\mathbf{p}_{\mathbf{r}}\cdot \left[
\mathbf{r}\times \left( 2\mathbf{S}+(1+k){\mbox{\boldmath
$\sigma$}}\right) \right]   \notag
\\
&&+\frac{G(2k-1)(\mathbf{p}_{\mathbf{r}}\cdot \mathbf{r)}}{2c^{2}\mu
r^{5}}\mathbf{p}_{\bm{\delta }}\cdot (\mathbf{r}\times
{\mbox{\boldmath
$\sigma$})}  \notag \\
&&-\frac{G(2k-1)}{2c^{2}r^{3}}\mathbf{p}_{\bm{\delta }}\cdot
(\mathbf{v}\times {\mbox{\boldmath
$\sigma$})}  \notag \\
&&-{\mbox{\boldmath $\delta$}}\mathbf{\cdot }\left(
\frac{\mathbf{p}_{ \mathbf{r}}}{\mu }-{\mbox{\boldmath
$\lambda$}}\right) {-}\frac{Gm}{r^{3}}
\mathbf{p}_{\bm{\lambda }}\cdot \mathbf{r}  \notag \\
&&\mathbf{-}{\frac{G}{c^{2}\mu r^{3}}}\mathbf{p}_{\bm{\lambda
}}\cdot \Biggl\{\mathbf{p}_{\mathbf{r}}\mathbf{\times
(}4\mathbf{S}+3{\mbox{\boldmath
$\sigma$})}  \notag \\
&&-\mathbf{r}\left[ (\mathbf{r}\times \mathbf{p}_{\mathbf{r}})\cdot
(2\mathbf{S}+(1+k)\mbox{\boldmath
$\sigma$})\right]   \notag \\
&&-{\frac{3(\mathbf{r}\cdot
\mathbf{p}_{\mathbf{r}}\mathbf{)}}{r^{2}}}\mathbf{r}\times
(2\mathbf{S}+(2-k){\mbox{\boldmath $\sigma$}})\Biggr\}.
\end{eqnarray}
where we added an extra term
$(2k-1)/(2c^{2}m)\mathbf{p}_{\mathbf{\delta }}\cdot
({\mbox{\boldmath ${\ddot{\lambda}}$}}\times {\mbox{\boldmath
$\sigma$})}$ because the $\mathbf{p}_{\mathbf{\delta }}\ $is
vanishing on the constraint surface, and we replaced the variables
${\mbox{\boldmath ${\dot{\lambda}}$},\mbox{\boldmath
${\ddot{\lambda}}$}}$ by the acceleration with
$\mathbf{p}_{\mathbf{r}}$ and $\mathbf{r}$ in Eq. (\ref{acc}) Thus,
the Hamilton's equations are consistent with the Euler-Lagrange
equations in Eqs. (\ref{ELcon}), and they are satisfied up to the SO
order as
\begin{eqnarray}
\mathbf{\dot{p}}_{\mathbf{r}} &=&-\frac{\partial
\mathcal{H}}{\partial \mathbf{r}},\qquad
\mathbf{\dot{p}}_{\bm{\lambda }}=-\frac{\partial
\mathcal{H}}{\partial {\mbox{\boldmath $\lambda$}}},\qquad
\mathbf{\dot{p}}_{\bm{\delta }}=-\frac{\partial
\mathcal{H}}{\partial {\mbox{\boldmath
$\delta$}}},  \notag \\
\mathbf{\dot{r}} &=&\frac{\partial \mathcal{H}}{\partial
\mathbf{p}_{\mathbf{r}}},\qquad {\mbox{\boldmath
${\dot{\lambda}}$}}=\frac{\partial \mathcal{H}}{\partial
\mathbf{p}_{\bm{\lambda }}},\qquad {\mbox{\boldmath
${\dot{\delta}}$}}=\frac{\partial \mathcal{H}}{\partial
\mathbf{p}_{\bm{\delta }}}.
\end{eqnarray}

\subsection{Appendix B: The elimination of acceleration: \textit{The method of the double zero}}

Barker and O'Connell proposed a procedure for the perturbation
method in which the acceleration terms can be eliminated from the
Lagrangian, which is called \textit{the method of the double zero}.
In this method the Lagrangian contains some lower-order conserved
quantities \cite{BOC80,BOC80b,BOC80c}. We are following this method.
Let's Lagrangian, Eq. (\ref{LagrangeSO}), can be written as
\begin{equation}
\mathcal{L}=\mathcal{L}_{N}+\mathcal{L}_{non\,\mathbf{a}}^{\prime
}+\mathcal{L}_{\mathbf{a}}^{\prime },  \label{Lagr}
\end{equation}
with
\begin{eqnarray}
\mathcal{L}_{N} &=&\frac{\mu }{2}\mathbf{v}^{2}+\frac{Gm\mu }{r},
\label{Newtonian} \\
\mathcal{L}_{non\,\mathbf{a}}^{\prime } &=&\frac{G\mu
}{2c^{2}r^{3}}\mathbf{v\cdot }\left[ \mathbf{r}\times \left(
4\mathbf{S}+3{\mbox{\boldmath
$\sigma$}}\right) \right] ,  \label{Lag_nona} \\
\mathcal{L}_{\mathbf{a}}^{\prime }
&=&-\frac{(2k-1)}{2c^{2}m}{\mbox{\boldmath $\sigma$}}\cdot \left[
\left( \mathbf{a+}\frac{Gm}{r^{3}}\mathbf{r}\right) \times \mu
\mathbf{v}\right] .  \label{Lag_a}
\end{eqnarray}
It can be seen that $\mathcal{L}_{non\,\mathbf{a}}^{\prime }$ does
not depend on SSC if we use the Newtonian-order acceleration
Lagrangian of Eq. (\ref{LagrangeSO}). Then we get the
$\mathcal{L}_{non\,\mathbf{a}}^{\prime }$. Our aim is to eliminate
the acceleration from $\mathcal{L}_{\mathbf{a}}^{\prime }$. Let's
consider the next double zero term
\begin{equation}
ZZ_{N}=-\frac{(2k-1)}{2c^{2}m}\left( {\mbox{\boldmath
$\sigma$}}_{0}{-\mbox{\boldmath $\sigma$}}\right) \cdot \left[
\left( \mathbf{a+}\frac{Gm}{r^{3}}\mathbf{r}\right) \times \mu
\mathbf{v}\right] .
\end{equation}
Here the Newtonian-order equation of motion is
$\mathbf{a}_{N}=-(Gmr^{-3})\mathbf{r}$, and  ${\mbox{\boldmath
${\dot{\sigma}}$}=0}$ so ${\mbox{\boldmath $\sigma$}}_{0}$ is the
conserved quantity up to SO order. Then
\begin{equation}
\mathcal{L}_{\mathbf{a}}^{\prime
}+ZZ_{N}=-\frac{(2k-1)}{2c^{2}m}{\mbox{\boldmath $\sigma$}}_{0}\cdot
\left[ \left( \mathbf{a+}\frac{Gm}{r^{3}}\mathbf{r}\right) \times
\mu \mathbf{v}\right] .
\end{equation}
Moreover, we note that the $\mathbf{S}$ and ${\mbox{\boldmath
$\sigma$}}$ spinvectors are conserved quantities in Eq.
(\ref{Lagr}), and we just follow the original paper of \cite{BOC86}.
Afterwards we consider the other double zero and \textit{total
time-derivative} terms
\begin{eqnarray}
\mathbf{ZZ} &=&\frac{2}{r^{2}}\left[ \left( \mathbf{a+}\frac{Gm}{r^{3}}\mathbf{r}
\right) \cdot \mathbf{r}\right] \left( \mathbf{L}_{0}{-}\mathbf{L}\right)  \notag \\
&&-\frac{1}{r^{2}}\left[ \left(
\mathbf{a+}\frac{Gm}{r^{3}}\mathbf{r}\right)
\cdot \left( \mathbf{L}_{0}{-}\mathbf{L}\right) \right] \mathbf{r,} \\
\mathbf{TTD} &=&\frac{d}{dt}\left( \frac{(\mathbf{v}\cdot
\mathbf{r)}}{r^{2}}\mathbf{L-}\frac{2(\mathbf{v}\cdot
\mathbf{r)}}{r^{2}}\mathbf{L}_{0}+\frac{(\mathbf{v}\cdot
\mathbf{L}_{0})}{r^{2}}\mathbf{r}\right) .
\end{eqnarray}
Here $\mathbf{L=}\mu \mathbf{r}\times \mathbf{v}$ is the angular
momentum vector and $\mathbf{L}_{0}$ is the conserved quantity for the
Newtonian-order. This distinction is important even in the lowest
order, because it is essential for the extraction of equations of
motion from the Lagrangian. We define the new Lagrangian which does not
contain the acceleration-dependent terms
\begin{eqnarray}
\mathcal{L}^{\prime \prime } &=&\mathcal{L}_{N}+\mathcal{L}_{non\,
\mathbf{a}}^{\prime }+\mathcal{L}_{\mathbf{a}}^{\prime }+ZZ_{N}  \notag \\
&&-\frac{(2k-1)}{2c^{2}m}{\mbox{\boldmath $\sigma$}}_{0}\cdot
(\mathbf{ZZ+TTD}),
\end{eqnarray}
so we get
\begin{eqnarray}
\mathcal{L}^{\prime \prime } &=&\frac{\mu
}{2}\mathbf{v}^{2}+\frac{Gm\mu }{r}+\frac{G\mu
}{2c^{2}r^{3}}\mathbf{v\cdot }\left[ \mathbf{r}\times \left(
4\mathbf{S}_{0}+3{\mbox{\boldmath
$\sigma$}}_{0}\right) \right]  \notag \\
&&-\frac{(2k-1)}{2c^{2}m}\Biggl\{\left( \frac{\mathbf{v}^{2}}
{r^{2}}-\frac{Gm}{r^{3}}-\frac{2(\mathbf{v}\cdot \mathbf{r})^{2}}
{r^{4}}\right) {\mbox{\boldmath $\sigma$}}_{0}\cdot (\mathbf{L}{-2}\mathbf{L}_{0})  \notag \\
&&-\left( \frac{Gm}{r^{5}}(\mathbf{r}\cdot \mathbf{L}_{0})+
\frac{2(\mathbf{v}\cdot \mathbf{r})}{r^{4}}(\mathbf{v}\cdot
\mathbf{L}_{0})\right) ({\mbox{\boldmath $\sigma$}}_{0}\cdot \mathbf{r)}  \notag \\
&&+\frac{(\mathbf{v}\cdot \mathbf{L}_{0})}{r^{2}}({\mbox{\boldmath
$\sigma$}}_{0}\cdot \mathbf{v)}\Biggr\}.
\end{eqnarray}
We have replaced the spinvectors $\mathbf{S}$ and ${\mbox{\boldmath
$\sigma$}}$ to $\mathbf{S}_{0}$ and ${\mbox{\boldmath
$\sigma$}}_{0}$ because these are conserved quantities in
$\mathcal{L}_{non\,\mathbf{a}}^{\prime }$. The equations of motion
can be derived from the acceleration-independent Lagrangian
$\mathcal{L}^{\prime \prime }$ with the replacement of
$\mathbf{S}_{0},$ ${\mbox{\boldmath $\sigma$}}_{0}$ and
$\mathbf{L}_{0}$ in the equations of motion by $\mathbf{S},$
${\mbox{\boldmath $\sigma$}}$ and $\mathbf{L}$, respectively.

\subsection{Appendix C: The Hamiltonian formalism for SSC II}

Let's consider the Hamiltonian formalism for SSC II,
\begin{eqnarray}
\mathcal{H} &=&\frac{\mathbf{p}^{2}}{2\mu }-\frac{G\mu m}{r}  \notag \\
&&+\frac{G}{2c^{2}r^{3}}\mathbf{r}\cdot \left[ \mathbf{p}\times
\left( 4\mathbf{S}+3{\mbox{\boldmath $\sigma$}}\right) \right],
\end{eqnarray}
where the limit of $k=1/2$ is not appropriate. The required limit is
$\mathbf{q}\rightarrow 0$ because the higher-order terms have to
disappear in this case. Then the usual Hamilton's equations are
\begin{eqnarray}
\mathbf{\dot{p}} &=&-\frac{G\mu
m}{r^{3}}\mathbf{r}-\frac{G}{2c^{2}r^{3}}\mathbf{p}\times \left(
4\mathbf{S}+3{\mbox{\boldmath
$\sigma$}}\right)  \notag \\
&&+\frac{3G}{2c^{2}r^{5}}\mathbf{r}\left[ (\mathbf{r}\times \mathbf{p})\cdot
(4\mathbf{S}+3\mbox{\boldmath $\sigma$})\right] ,  \label{Ham_p} \\
\mathbf{\dot{r}} &\mathbf{=}&\frac{\mathbf{p}}{\mu
}-\frac{G}{2c^{2}r^{3}}\mathbf{r}\times \left(
4\mathbf{S}+3{\mbox{\boldmath $\sigma$}}\right) .  \label{Ham_r}
\end{eqnarray}
It is interesting to note that the total angular momentum has a
simple form, $\mathbf{L=r}\times \mathbf{p}$, but if we use Eq.
(\ref{Ham_r}), we get the complicated form of Eq. (\ref{Lageq}) in
the Lagrangian formalism for SSC II, as
\begin{equation}
\mathbf{L=}\mu \mathbf{r}\times \mathbf{v}+\frac{G\mu
}{2c^{2}r^{3}}\mathbf{r}\times \left[ \mathbf{r}\times \left(
4\mathbf{S}+3{\mbox{\boldmath $\sigma$}}\right) \right] .
\end{equation}
We assume that the canonical momentum has the decomposition
$\mathbf{p=}p_{r}\mathbf{e}_{r}+p_{\theta }\mathbf{e}_{\theta
}+p_{\phi }\mathbf{e}_{\phi }$ with orthonormal basis
$(\mathbf{e}_{r}\mathbf{,e}_{\theta }\mathbf{,e}_{\phi })$ in an
inertial frame fixed by the conserved total angular momentum vector
$\mathbf{J}$. We use the decomposition of
$\mathbf{v}=v_{r}\mathbf{e}_{r}+v_{\theta }\mathbf{e}_{\theta
}+v_{\phi }\mathbf{e}_{\phi }$ from the simple definition of
$\mathbf{\hat{r}}\equiv \mathbf{e}_{r}$. In Eq. (\ref{Ham_r})
\begin{equation}
\mathbf{p}^{2}\mathbf{=}p_{r}^{2}+p_{\theta }^{2}+p_{\phi
}^{2}=p_{r}^{2}+\frac{L^{2}}{r^{2}},  \label{ppp}
\end{equation}
we have used the identity
$\mathbf{p}^{2}=(\mathbf{\mathbf{\hat{r}}\cdot
p)}^{2}\mathbf{+(\mathbf{\hat{r}}\times p)}^{2}$. Here we can
rewrite the simple relationship
$\mathbf{v=}\dot{r}\mathbf{e}_{r}\mathbf{\mathbf{+}}r\dot{\theta}\mathbf{e}_{\theta
}+r\dot{\phi}\sin \theta \mathbf{e}_{\phi }$. Then the radial
equation is ($p_{r}^{2}=\mu ^{2}\dot{r}^{2}$)
\begin{eqnarray}
\dot{r}^{2} &=&\frac{2E}{\mu }+\frac{2Gm}{r}-\frac{L^{2}}{\mu ^{2}r^{2}}
\notag \\
&&-\frac{G\left( 4\mathbf{S}+3{\mbox{\boldmath
$\sigma$}}\right) \cdot \mathbf{L}}{c^{2}\mu r^{3}},
\end{eqnarray}
which is the same as Eq. (\ref{radial}). Let us consider the angular
motion. We compute the quantity $\mathbf{L\cdot e}_{\mathbf{z}}$
(where the unit vector $\mathbf{e}_{\mathbf{z}}$ is
$\mathbf{e}_{\mathbf{z}}=\mathbf{e}_{r}\cos \theta
-\mathbf{e}_{\theta }\sin \theta $ in spherical polar coordinates),
so $\mathbf{L\cdot e}_{\mathbf{Z}}\equiv L\cos \Theta =p_{\phi
}r\sin \theta $. Using Eq. (\ref{ppp}) we get the components of
$\mathbf{p}$, where $\Theta $ is the angle between $\mathbf{L}$ and
$\mathbf{J}$.
\begin{eqnarray}
p_{\phi } &=&\frac{L\cos \Theta }{r\sin \theta }, \\
p_{\theta }^{2} &=&\frac{L^{2}}{r^{2}}\left( 1-\frac{\cos ^{2}\Theta }{\sin
^{2}\theta }\right) .
\end{eqnarray}
Using the equations $r\sin \theta \dot{\phi}=\mathbf{e}_{\phi }\cdot
\mathbf{v}$, $r\dot{\theta}=\mathbf{e}_{\theta }\cdot \mathbf{v}$
and the Hamilton Eq. (\ref{Ham_r}), we get ($\mathbf{\hat{r}}\times
\mathbf{e}_{\theta }=\mathbf{e}_{\phi }$, $\mathbf{\hat{r}}\times
\mathbf{e}_{\phi }=-\mathbf{e}_{\theta }$)
\begin{eqnarray}
\dot{\phi} &=&\frac{L\cos \Theta }{\mu r^{2}\sin ^{2}\theta }\left(
1-\frac{G\mu \sin \theta S_{\theta }}{2c^{2}Lr\cos \Theta }\right) ,
\label{polarangle1} \\
\dot{\theta}^{2} &=&\frac{L^{2}\left( 1-\frac{\cos ^{2}\Theta }{\sin
^{2}\theta }\right) }{\mu ^{2}r^{4}}\left( 1+\frac{G\mu S_{\phi
}}{c^{2}Lr\sqrt{1-\frac{\cos ^{2}\Theta }{\sin ^{2}\theta }}}\right)
, \label{polarangle2}
\end{eqnarray}
where the $S_{\theta }\equiv \mathbf{e}_{\theta }\cdot \left(
4\mathbf{S}+3{\mbox{\boldmath $\sigma$}}\right) $ and the $S_{\phi
}\equiv \mathbf{e}_{\phi }\cdot \left( 4\mathbf{S}+3{\mbox{\boldmath
$\sigma$}}\right) $ are shorthand notations. The equations for polar
angles $\theta $ and $\phi $ can be transformed in Euler angles
equations ($\varphi ,\Upsilon ,\Theta $) if we write the unit
separation vector $\mathbf{\hat{r}} $ of Descartes components in an
invariant system fixed to $\mathbf{J}$, as in
\begin{eqnarray}
\cos \theta &=&\sin \varphi \sin \Theta ,  \notag \\
\sin (\phi -\Upsilon )\sin \theta &=&\sin \varphi \cos \Theta ,  \notag \\
\cos (\phi -\Upsilon )\sin \theta &=&\cos \varphi .  \label{Euler3}
\end{eqnarray}
These transformation identities are the same as the other 3 angles
$\varphi ,\Upsilon ,\Theta _{N}$, but we use the substitution of
$\Theta \rightarrow \Theta _{N}$. The relationship between the two
angles from the components of $\mathbf{L}_{\mathbf{N}}=L\left[
1-\lambda _{so}/(2L^{2})\right] (0,-\cos \Theta _{N}/\sin \theta
,\sqrt{1-\cos ^{2}\Theta _{N}/\sin ^{2}\theta })$ and
$\mathbf{L}=L(0,-\cos \Theta /\sin \theta ,\sqrt{1-\cos ^{2}\Theta
/\sin ^{2}\theta })$ in a spherical coordinate system is
\begin{eqnarray}
\cos \Theta &=&\cos \Theta _{N}\biggl [1+\frac{G{\mu }\sqrt{1-\frac{\cos
^{2}\Theta _{N}}{\sin ^{2}\theta }}}{2c^{2}rL}  \notag \\
&&\times \left( \frac{\sin \theta }{\cos \Theta _{N}}\sqrt{1-\frac{\cos
^{2}\Theta _{N}}{\sin ^{2}\theta }}S_{\theta }+S_{\phi }\right) \biggr ],
\label{inclination_trans}
\end{eqnarray}
where we used the quantity $\lambda _{so}=2\mathbf{L\cdot L}_{SO}$ in
a spherical coordinate system, as in
\begin{equation}
\lambda _{so}=\frac{G\mu L}{c^{2}r}\left( \frac{\cos \Theta
_{N}}{\sin \theta }S_{\theta }-\sqrt{1-\frac{\cos ^{2}\Theta
_{N}}{\sin ^{2}\theta }}S_{\phi }\right),
\end{equation}
where the inclination angles can be replaced by $\Theta
_{N}\leftrightarrow \Theta $ because these angles appear in
leading-order contributions. The time evolution from Eqs.
(\ref{Euler3}) is
\begin{equation}
\dot{\theta}=-\sin (\phi -\Upsilon )\dot{\Theta}-\sqrt{1-\frac{\cos
^{2}\Theta }{\sin ^{2}\theta }}\dot{\varphi}.  \label{angleidentity}
\end{equation}
After eliminating $\dot{\theta}$ we get the final formula for the evolution
of $\varphi $
\begin{equation}
\dot{\varphi}=\frac{L}{\mu r^{2}}\left( 1+\frac{G\mu S_{\phi
}}{2c^{2}Lr\sqrt{1-\frac{\cos ^{2}\Theta }{\sin ^{2}\theta
}}}\right) -\dot{\Upsilon}\cos \Theta ,  \label{Hamilton_angle}
\end{equation}
which agrees with the Eq. (6.13) in \cite{Tessmer} but we chose the
sign $-$ after the extraction of Eq. (\ref{polarangle2}). The last
term is $\dot{\Theta}\sin (\phi -\Upsilon )(1-\cos ^{2}\Theta /\sin
^{2}\theta )^{-1/2}$, which corresponds to the equation for
$\dot{\Upsilon}$ in Eq. (\ref{angle2}) interchanging $\Theta
_{N}\leftrightarrow \Theta $ because quantity $\dot{\Theta}$ (and
$\dot{\Theta}_{N}$) has a linear order $\mathcal{O}(c^{-2})$, and
here we can use the approximation $\Theta $ $\approx \Theta _{N}$
(see Eqs. (\ref{angle3}), (\ref{angle3b})). If we assume the equal
masses of single spin cases, then $\Theta $ is a constant and the
second term is zero in Eq. (\ref{Hamilton_angle}). Since
$\mathbf{J}$ is conserved using the Eq. (\ref{angleidentity}), we
get $\mathbf{e}_{\phi }\cdot $ $\mathbf{S=}$ \thinspace $-$ $\mu
r^{2}\dot{\theta}+\mathcal{O}(c^{-2})$. Then we get a similar
angular equation Eq. (4.29) as in \cite{Gopa}. In general cases if
we compute the $S_{\phi }$ (and $S_{\theta }$) up to the
leading-order for Eq. (\ref{Hamilton_angle}), we get the same
angular Eqs. (\ref{angle1}) and (\ref{angle2}). In scalar products
$\Theta _{N}\leftrightarrow \Theta $ are interchangeable, which
correspond to the equations
\begin{eqnarray}
S_{\phi } &=&\sqrt{1-\frac{\cos ^{2}\Theta }{\sin ^{2}\theta }}\frac{
\left( 4\mathbf{S}+3{\mbox{\boldmath $\sigma$}}\right) \cdot \mathbf{L}}{{L}}, \\
S_{\theta } &=&-\frac{\cos \Theta }{\sin \theta }\frac{\left(
4\mathbf{S}+3{\mbox{\boldmath $\sigma$}}\right) \cdot
\mathbf{L}}{L}.
\end{eqnarray}
If we want to compare the angular Eqs. (\ref{polarangle1}) and
(\ref{polarangle2}) to our earlier results in \cite{Kepler}, we need
to use the transformation between \ $\Theta $ and $\Theta _{N}$ in
Eq. (\ref{inclination_trans}).

\end{document}